\documentclass[12pt,a4paper]{article}
\pdfoutput=1
\usepackage{jheppub}
\usepackage[FIGTOPCAP]{subfigure}
\usepackage{amsmath}
\usepackage{amsfonts}
\usepackage{amssymb}
\usepackage{tensor}
\usepackage{amsmath}
\usepackage{amssymb}
\usepackage{amsthm}
\usepackage{psfrag}
\usepackage{graphicx}
\usepackage{color}
\usepackage{bbm}
\usepackage{color}
\usepackage{hyperref}

\newcommand{\exclude}[1]{}

\newcommand{\beq}{\begin{equation}}
\newcommand{\eeq}{\end{equation}}
\newcommand{\bea}{\begin{eqnarray}}
\newcommand{\eea}{\end{eqnarray}}
\pdfoutput=1
\long\def\/*#1*/{}

\newcommand{\p}{\partial}

\newcommand{\OO}{{\cal O}}

\newcommand{\junk}[1]{}

\title{\center{Phases of holographic superconductors with broken translational symmetry}}
\author[a]{Matteo Baggioli, \note{mbaggioli@ifae.es}}
\author[b]{Mikhail Goykhman \note{goykhman89@gmail.com}}
\affiliation[a]{Departament de F\'{i}sica and IFAE, Universitat Aut\`{o}noma de Barcelona, \\ Bellaterra, 08193, Barcelona, Spain.}
\affiliation[b]{Enrico Fermi Institute, University of Chicago,\\
5620 S. Ellis Av., Chicago, IL 60637, USA}

\abstract{
We consider holographic superconductors in a broad class of massive gravity backgrounds.
These theories provide a holographic description of a superconductor with broken translational
symmetry.
 Such models exhibit a rich phase structure:
depending on the values of the temperature and the disorder strength
the boundary system can be in superconducting, normal
metallic or normal pseudo-insulating phases.
Furthermore the system supports interesting collective excitation of the charge 
carriers, which appears in the normal phase, persists in the superconducting
phase, but eventually gets destroyed by the superconducting condensate.
We also show the possibility of
building a phase diagram of a system with the superconducting
phase occupying a dome-shaped region on the temperature-disorder plane.}

\begin{document}

\maketitle
\color{black}
\section{Introduction}

Holographic superconductors have recently received
a new wave of attention.
It originated from several attempts \cite{Zeng:2014uoa,Ling:2014laa,Andrade:2014xca,Kim:2015dna,Erdmenger:2015qqa,Arean:2014oaa,Horowitz:2013jaa} to provide a holographic
description of systems which resemble more of the real-world
superconductors. One of the essential features of the original holographic superconductor
proposal of \cite{Hartnoll:2008vx,Hartnoll:2008kx} is that it describes the system which exists in
two states: a superconducting state which has a non-vanishing charge condensate,
and a normal state which is a perfect conductor. As a direct conseguence, already in the normal phase the static electric response, namely the DC conductivity ($\omega=0$), is infinite. This is a straightforward consequence of the translational
invariance of the boundary field theory, which leads to the fact that the charge
carriers do not dissipate their momentum, and accelerate freely
under an applied external electric field.
Therefore
one is motivated to introduce momentum dissipation into the holographic framework, breaking the translational invariance of the dual field theory.
It is definitely interesting 
to construct a holographic
superconductor on top of such dissipative backgrounds which is indeed going to have a finite DC conductivity in the normal phase, clearly distinguishable from the infinite one in the superconducting phase.\\

One efficient method to implement such a feature relies on the possibility of breaking diffeomorphism invariance in the bulk via giving the graviton a mass, as it has been proposed in \cite{Vegh:2013sk}. It is very convenient to recast these Lorentz symmetry violating massive gravity theories into a covariant form introducing the Stueckelberg fields, namely the extra degrees of freedom appearing as a consequence of breaking of the diffeomorphism symmetry (see \cite{Rubakov:2008nh} for more details).\\

In the context of applied holography this construction was analyzed for the first time in \cite{Andrade:2013gsa} where momentum dissipation
in the field theory was achieved by switching on neutral scalar operators depending linearly
on the spatial coordinates of the boundary.
These scalar fields on the boundary source the
neutral scalar fields in the bulk. The resulting bulk
system describes a holographic dual of the field theory with broken translational symmetry.
Such a system possesses a finite DC conductivity~\cite{Andrade:2013gsa}\footnote
{See \cite{Davison:2013jba,Blake:2013owa,Blake:2013bqa,Davison:2013txa}
for further studies about Massive Gravity as an effective description for Momentum Dissipation.}.\\

The original idea of \cite{Andrade:2013gsa} has been put in a broader context in \cite{Baggioli:2014roa},
where the most general form for the
Lagrangian of the neutral scalars has been introduced\footnote{A more restrictive generalization has been analyzed in \cite{Taylor:2014tka}.}.
This Lagrangian is weakly constrained by the consistency conditions in the bulk, which avoid ghost excitations and gradient instabilities \cite{Baggioli:2014roa}.
It turns out that imposing physical consistency of the theory still leaves enough freedom to construct models,
which exhibit new non-trivial features.\\

To be more specific, one can build models which possess the following attractive properties.
The first one is an increase of conductivity as a function of temperature, for temperatures lower than a
certain critical value $T_0$,
\beq
\frac{d\sigma_{DC}(T)}{dT}>0\,,\qquad T<T_0\,.\label{pseudoins}
\eeq
This property bears a resemblance to an insulating behavior, with the
population of the conducting energy band depleting upon lowering the temperature.
Still, it awaits a better understanding, because of an essentially non-vanishing
value of the DC conductivity at zero temperature. We refer to the
state (\ref{pseudoins}) as {\it pseudo-insulating}.
The second new feature of the model
is an appearance of an extra structure in the optical conductivity.
For temperatures lower than a certain critical value $T'$, there appears a peak
in the optical conductivity, signaling a new long-lived collective propagating excitation of the charge carriers\footnote{It is really tempting to make a comparison to polaron physics, see, {e.g.}, \cite{Klimin:2014}.}.\\

This paper is based on the idea to generalize
the construction of \cite{Kim:2015dna,Andrade:2014xca}
to the more generic effective models for momentum dissipative systems, proposed in \cite{Baggioli:2014roa}.
The main questions which we aim to answer are the following:\\

1. Can one construct a model of holographic superconductor
which is separated by the lines of the second order phase transition
from the normal metallic phase and the normal pseudo-insulating phase (\ref{pseudoins})?\\

2. Does the peak in the optical conductivity of \cite{Baggioli:2014roa}
continue to exist in the superconducting phase\footnote{See also \cite{Horowitz:2008bn}, where non-trivial structure has been observed in
the optical conductivity of a holographic superconductor.}?\\

We have found that the answers
are:\\

1. Yes, by combining the idea of \cite{Baggioli:2014roa} with the setting
of a holographic superconductor one can obtain a system
with a rich phase diagram where three different phases are present:
superconductor, metal, and pseudo-insulator.
\\

2. The peak in optical conductivity continues to exist
in the superconducting phase, as the temperature
is lowered below a critical temperature $T_c$
of the superconducting
phase transition. However, at a certain temperature $T=T''$ the peak disappears.
\\

\begin{figure}
\begin{center}
\includegraphics[width=.55\textwidth]{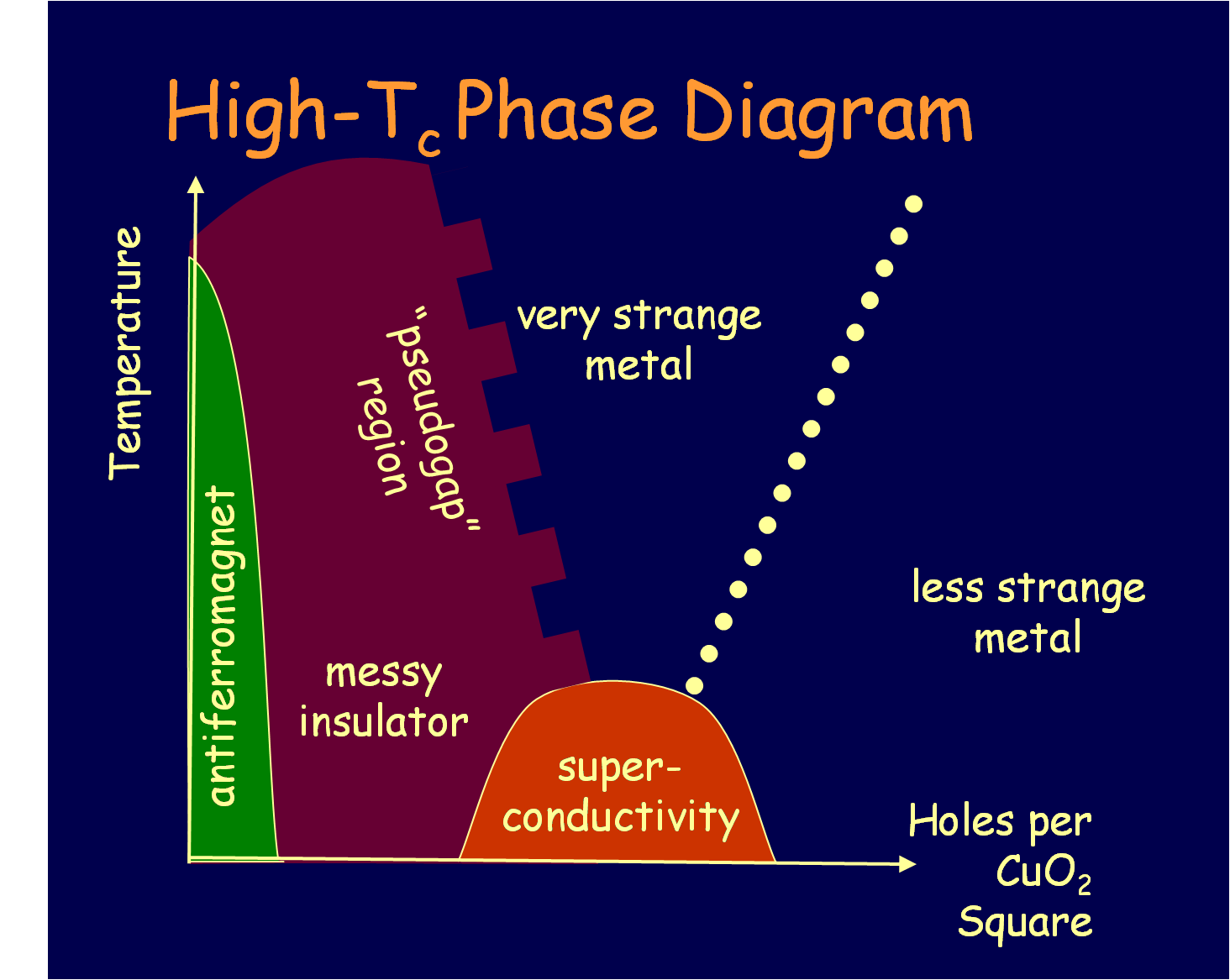}
\end{center}
\caption{Schematic phase diagram of a real cuprate high-Tc superconductor.}
\label{fig:DomeIntro}
\end{figure}

Furthermore, we attempt to construct a holographic model with the phase diagram containing a superconducting state inside a dome-shaped region. (see, e.g., \cite{Lee:2004} for the discussion about dome phase diagrams in condensed matter theory and figure \ref{fig:DomeIntro} for a sketch of the realistic situation\footnote{Note that in nature the axes are temperature and doping while in our case doping is replaced by disorder-strength; in this sense we do not aim to construct an holographic dual of a real dome-shaped phase diagram typical of high-Tc superconductors.})\color{black}.
The most successful result would be to have a superconducting dome,
separated from an insulating normal state at smaller values of the disorder strength parameter\color{black}, and a metal normal phase at its larger values.
In figure \ref{fig:DomeIntro} we provide a schematic sketch of what we would like to approach,  the phase diagram for High-Tc superconductors.\color{black}\\

We will demonstrate that implementing the momentum dissipation models of \cite{Baggioli:2014roa}
in the holographic superconductor framework can indeed lead to a superconducting
dome, located between pseudo-insulating and metallic phases. However, it appears that such models
are too restricted to describe superconducting dome with realistic critical temperature
of the superconducting phase transition. We have found that the critical temperature of
the dome $T_c(\alpha)$, where $\alpha$ is the magnitude of the
translational symmetry breaking, is bounded from above by a small number (in units of charge density),
of the order of $10^{-8}$. This makes the numerical calculation at finite temperature hopeless.\\

Nevertheless, at zero temperature it is possible to have analytical control of the SC instability through
the BF bound reasonings and
show the existence of a superconducting dome. \color{black}In the Discussion section \ref{section7} we provide
a few ideas to generalize our model, which might be useful to obtain
a superconducting dome with a reasonably higher values of the critical temperature.
We will be considering charged black brane
backgrounds with the neutral scalar fields having vacuum profiles, depending linearly on the spatial
coordinates:
\beq
\phi^x=\alpha \,x\,,\qquad \phi^y=\alpha\,y\,.\label{phixy}
\eeq
This configuration (\ref{phixy}) breaks translational symmetry (and Lorentz invariance) of the boundary field theory but keeps untouched energy conservation. Within this choice we are going to retain homogeneity and rotational invariance. It would be interesting to reproduce the same sort of computations in an anisotropic setup as in \cite{Erdmenger:2015qqa}.
Besides the parameter $\alpha$, describing the magnitude of the translational
symmetry breaking, we will also introduce another parameter $m$,
which will be primarily important in the models with non-linear
action for the neutral scalars (\ref{phixy}).
\\

We consider the system at a finite charge
density, which corresponds holographically to the time-like component
of the $U(1)$ gauge field
having a non-trivial radial profile in the bulk, $A_t (u)$.
The charged scalar $\psi$ is dual to the condensate ${\cal O}$
of charge carriers. When the v.e.v. of the condensate is non-vanishing, $\langle {\cal O}\rangle\neq 0$,
the system is in a superconducting phase. This corresponds holographically to a
non-trivial configuration $\psi(u)$ in the bulk, with the vanishing
source coefficient of the near-boundary expansion of the $\psi(u)$ \cite{Hartnoll:2008vx,Hartnoll:2008kx}.\\

We will study various superconducting systems, distinguished by the choice of the Lagrangian $V(X)$ for the neutral scalar fields, where
\begin{equation}
X=\frac{1}{2}\,L^2\,g^{\mu\nu}\p_\mu\phi^I\p_\nu\phi^I\,,
\end{equation}
and $L$ is the radius of AdS.
In this paper we will be mostly interested in the following models\footnote{In models \ref{model1} and \ref{model3} we introduced the prefactor of $1/m^2$ into the definition of $V(X)$. Such change of notation will render $\alpha$ to be the only translational symmetry breaking parameter in the models \ref{model1} and \ref{model3}. In model \ref{model2} instead  both $m$ and $\alpha$ are independent parameters and we decided to avoid any rescaling in the definition of $V(X)$.\color{black}}:
\begin{align}
{\bf model\;\; 1\;:}\qquad V(X)&=\frac{X}{2\,m^2}\,,\label{model1}\\
{\bf model\;\; 2\;:}\qquad V(X)&=X+X^5\label{model2}\\
{\bf model\;\; 3\;:}\qquad V(X)&=\frac{X^N}{2\,m^2}\,,\qquad N\neq 1\label{model3}
\end{align}
The model (\ref{model1}) 
gives the simplest way to describe the fields $\phi^I$ and has
been proposed in \cite{Andrade:2013gsa}.\\

We will argue that already in the simple case of (\ref{model1}) it is possible to have
a superconducting dome. We will demonstrate this analytically at zero temperature.
Interestingly, the dome
is achieved for the scaling dimension $\Delta$ and the charge $q$ of the scalar 
$\psi$,
restricted to the small vicinity of the ``dome" point, which we have found to be
\beq
(\Delta_d,\;q_d)=(2.74,\; 0.6)\,. \label{domecr}
\eeq
In this case the superconducting dome exists in the middle
of a normal metallic phase (the model (\ref{model1}) does not allow an insulating phase).
\\

We will show that the model with the non-linear Lagrangian (\ref{model2})
also possesses the superconducting dome near the point (\ref{domecr}).
In this case it is possible to engineer a model where the dome is separated
from metallic phase at larger values of the translational symmetry breaking parameter $m$,
and from a pseudo-insulating phase at smaller values of $m$. This situation is qualitatively the closest one to the actual real phase diagram for High-Tc superconductors. It is important no notice that our dome is constructed dialing the disorder-strentgh parameter of the theory, while the actual dome in High-Tc Superconductors depends on the doping\footnote{If our parameter was the doping of the material it would affect the charge density of the boundary theory and this does not happen in our model.\color{black}} of the material. We are not aware of experimental phase diagrams where the SC dome occurs as function of increasing disorder-strength.\color{black}\\

To support our statement about the superconducting dome with such a  small critical
temperature $T_c$, we will calculate numerically the dependence
of the critical temperature for the models (\ref{model1}), (\ref{model2}),
on the scaling dimension $\Delta$ and the charge $q$.
We will show that as the $(\Delta,\; q)$ approach the dome point (\ref{domecr}),
the critical temperature quickly declines.\\

The rest of this paper is organized as follows.
In the next Section~\ref{section1} we set up the model
which we will be studying in this paper. We consider the general Lagrangian $V(X)$
for the massless neutral scalar fields. In Section~\ref{section2} we review the properties of the normal phase solution. In Section~\ref{section3} we study the conditions for its instability towards formation of a non-trivial
profile of scalar hair. From the field theory point of view this corresponds
to a superconducting phase transition. In Section~\ref{section4} we focus on the features of the broken phase, the
condensate and the grand potential,
demonstrating explicitly the second order phase transition at $T=T_c$. In Section~\ref{section5}
we study the optical conductivity in the normal and superconducting phases.
In Section~\ref{section6} we describe the way to construct a superconducting dome
in the middle of a metallic phase, for the model (\ref{model1}),
and between pseudo-insulating and metallic phases, for the (\ref{model2}).
We discuss our results in Section~\ref{section7}.
Appendix~\ref{appendix2} contains further details about the 
calculations of the condensate and the grand potential.
Appendix~\ref{appendix1} is dedicated
to derivation of the on-shell action for bulk fluctuations, which are
holographically dual to current and momentum operators on the boundary.

\section{Setting up the model}\label{section1}

In this section we introduce the model, which we will be studying in this paper.
We begin by writing down the action and equations of motion of the bulk theory.
We proceed by deriving
equations of motion for the general ansatz, describing the charged black brane geometry,
with linearly-dependent sources for $\phi^I$, and radially dependent charged scalar $\psi(u)$.
Then we will review the normal-state solution of the model, which has a trivial charged scalar field
profile $\psi\equiv 0$.

\subsection{Action and equations of motion}

The total action of our model is :
\beq
I=I_1+I_2+I_3\,,\label{Itotb}
\eeq
where we have denoted the Einstein-Maxwell terms $I_1$,
the neutral scalar terms $I_2$, and the charged scalar terms $I_3$;
\begin{align}
I_1&=\int d^{d+1}x\,\sqrt{-g}\,\left[R-2\Lambda-\frac{L^2}{4}F_{\mu\nu}F^{\mu\nu}\right]\,,\notag\\
I_2&=-2m^2\int d^{d+1}x\,\sqrt{-g}\,V\left(X\right)\,,\label{Iact}\\
I_3&=-\int d^{d+1}x\,\sqrt{-g}\,\left(|D\psi|^2+M^2|\psi|^2+\kappa\, H\left(X\right)\,|\psi|^2\right)\notag\,.\label{action}
\end{align}
We have inserted an additional coupling $m^2$ in front of the potential $V(X)$ which is going to be redundant for the monomial cases \ref{model1} and \ref{model3} where we decided in fact to reabsorb it into the definition of $V(X)$. In this way for those cases we are left with just one parameter $\alpha$ which is going to represent the disorder-strength in the system. In the case of the polinomial potential \ref{model2} $m^2$ is going to be an independent parameter in addition to $\alpha$. \color{black}
We have introduced an extra coupling $\kappa$, between the charged scalar $\psi$ and the 
neutral scalars $\phi^I$.
In this paper we will be mostly considering $\kappa=0$, and comment on the models
with non-vanishing $\kappa$ in the discussion section \ref{section7}.
We have defined
\begin{equation}
X=\frac{1}{2}\,L^2\,g^{\mu\nu}\p_\mu\phi^I\p_\nu\phi^I\,.
\end{equation}
We denote $D_\mu\psi =(\p_\mu-i\,q\,A_\mu)\psi$ to be the standard covariant derivative of the scalar $\psi$
with the charge $q$.
We fix the cosmological constant to be $\Lambda=-3/L^2$.
In this paper we will consider $4$-dimensional bulk, $d=3$.

The equations of motion following from the action $I$ read\footnote{When $V(X)=X/2m^2$, we recover the equations of \cite{Kim:2015dna}.} :
\begin{align}
&R_{\mu\nu}-\frac{1}{2}\left(R-2\Lambda-\frac{L^2}{4}F^2-|D\psi|^2-(M^2+\kappa\,H)|\psi|^2-2m^2V\right)g_{\mu\nu}\notag\\
&=\left(m^2\,\dot V+\frac{1}{2}\,\kappa \,\dot H\,|\psi|^2\right)\,\p_\mu\phi^I\p_\nu\phi^I+\frac{L^2}{2}F_{\mu\lambda}F_{\nu}^{\;\;\lambda}+
\frac{1}{2}(D_\mu\psi D_\nu\psi^\star +D_\nu\psi D_\mu\psi^\star)\notag\\
&\frac{1}{\sqrt{-g}}\,\p_\mu\left(\sqrt{-g}\, F^{\mu\nu}\right)-i\frac{q}{L^2}(\psi^\star D^\nu\psi-\psi D^\nu\psi^\star)=0\label{Einseq}\\
&\frac{1}{\sqrt{-g}}\,D_\mu \left(\sqrt{-g}\,D^\mu\psi\right)-(M^2+\kappa\, H)\psi=0\notag\\
&\p_\mu\left(\sqrt{-g}\,(2m^2\,\dot V+\kappa\,H\,|\psi|^2)\,g^{\mu\nu}\p_\nu\phi^I\right)=0\,,\notag
\end{align}
where the dot stands for a derivative w.r.t. $X$,
\beq
\dot V(X)\equiv \frac{dV}{dX}\,,\qquad \dot H(X)\equiv \frac{dH}{dX}\,.
\eeq

\subsection{Background}
We consider the following black brane ansatz for the background:
\begin{align}
&ds^2=L^2\left(-\frac{1}{u^2}f(u)e^{-\chi(u)}dt^2+\frac{1}{u^2}(dx^2+dy^2)+\frac{1}{u^2f(u)}du^2\right)\notag\\
&\phi^I=\alpha\, \delta^I_i\,x^i\,,\,\,\,\,\,\,\,\,\,\,\quad I,i=x,y\,.\label{ansb}\\
&A=A_t(u)du\,,\qquad \psi=\psi (u)\,.\notag
\end{align}
The $\phi^I$ scalars have profiles linear in the spatial coordinates $x,\,y$ of the boundary.
They effectively describe momentum dissipation mechanisms in the boundary field
theory, making the DC conductivity of the theory finite\footnote{These fields are dual to marginal scalar operators whose sources explicitly break translational symmetry. Exploiting the shift invariance for these operators it is possible to retain the homogeinity of the background such that the metric and the charged scalar (and as a conseguence the stress tensor and the SC order parameter) do not depend on the spatial coordinates at all.\color{black}} \cite{Andrade:2013gsa}.
We will take $\psi$ to be real-valued, since due to
the $u$ component of Maxwell equations the phase of the complex field $\psi$ is a constant.
We are looking for charged
black brane solutions with a scalar hair where $u_h$ is the position of the horizon, and the boundary
is located at  $u=0$.
We allow for non-trivial $\chi(u)$ because we want to have
in general a non-trivial $\psi(u)$. If $\psi =0$, then $\chi=0$.

The resulting equations of motion read:
\begin{align}
&\frac{q^2\,u\,e^\chi\, A_t^2\,\psi^2}{f^2}-\,\chi'+u\,\psi'^2=0\label{sEinst}\\
&\psi'^2-\frac{2\,f'}{u\,f}+\frac{e^\chi u^2 A_t'^2}{2 f}+\frac{M^2 L^2 \psi^2}{u^2 f}+\frac{\kappa\,L^2\, H \psi^2}{u^2 f}+\frac{e^\chi q^2 A_t^2 \psi^2}{f^2}\notag \\&+\frac{2 m^2 L^2 V}{u^2 f}+\frac{2\Lambda L^2}{u^2 f}+\frac{6}{u^2}\,=0\label{lEinst}\\
&\frac{2 q^2  A_t \psi^2}{u^2 f}-\frac{\chi'}{2}A_t'-A_t''=0\label{Maxb}\\
&\psi''+\left(-\frac{2}{u}+\frac{f'}{f}-\frac{\chi'}{2}\right)\psi'
+\left(\frac{e^\chi q^2 A_t^2}{f^2}-\frac{M^2 L^2}{u^2 f}-\frac{\kappa H\,L^2 }{u^2 f}\right)\psi\,=0\label{scfeq}
\end{align}
The Hawking temperature of the black brane (\ref{ansb}) is given by:
\beq
T\,=\,-\,\frac{f'(u_h)}{4\pi}\,e^{-\frac{\chi(u_h)}{2}}\,.\label{Tdef}
\eeq
Using eqs. (\ref{sEinst})-(\ref{scfeq}), the temperature can be written as:
\beq
T\,=\,-\,\frac{e^{-\frac{\chi}{2}}}{16\pi u_h}\left(-12+4 m^2 L^2V+2 (M^2+\kappa \,H)L^2 \psi^2+e^\chi u_h^4 A_t'^2\right) \,.\label{T}
\eeq
with all the fields evaluated at the horizon $u_h$.

\subsection{Normal phase}

In the case of a non-trivial condensate $\psi(u)$ it is in general impossible to solve the background
equations of motion (\ref{sEinst})-(\ref{scfeq}) analytically. However, when $\psi(u)=0$, the solution is known
\cite{Baggioli:2014roa}.

From now on we will fix the coupling $\kappa$ to zero,
\beq
\kappa=0\,.
\eeq
The resulting normal phase background is given by:
\begin{align}
& \psi(u)=0\,,\,\,\,\,\,\,\chi(u)=0\,\,,\\
& A_t(u)=\mu - u \rho\,,\qquad \mu=\rho\,u_h\,,\\
& f(u)= u^3 \int_{u_h}^{u} \left(-\frac{3}{y^4}+\frac{m^2 L^2 V(\alpha^2 y^2)}{y^4}+\frac{\rho^2}{4}\right)\,dy
\label{eomsnormalt}
\end{align}
Due to (\ref{T}) the temperature in the normal state reads:
\beq
\label{normalT}
T\,=\,-\,\frac{1}{16\pi u_h}\left(-12+4 m^2 L^2V+u_h^4 \rho^2\right) \,.
\eeq
All the features of this normal phase solution are going to be reviewed in detail in the following section.\\
\color{black}

\section{Normal phase features}\label{section2}

\begin{figure}
\centering
\includegraphics[width=.45\textwidth]{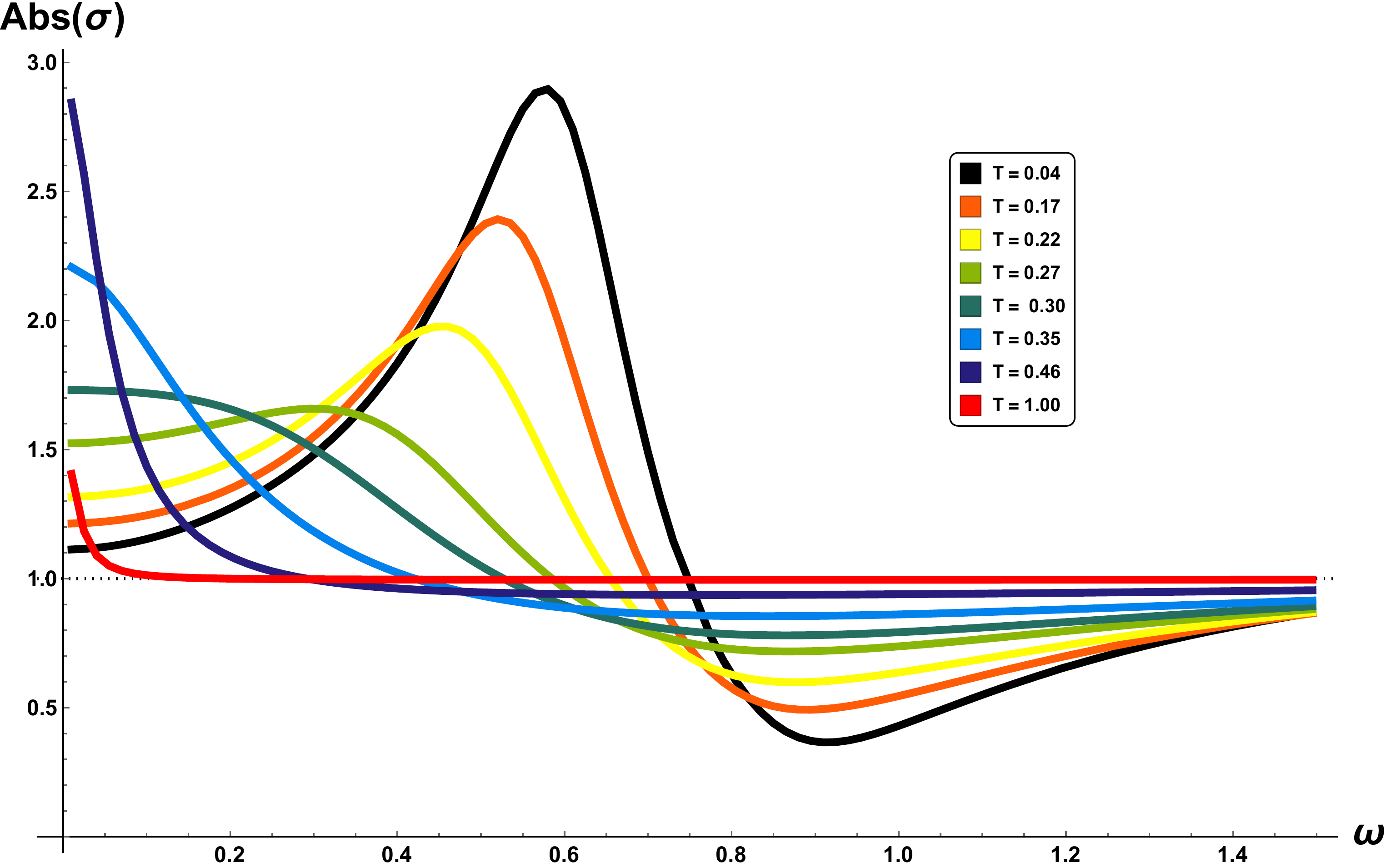}
\includegraphics[width=.38\textwidth]{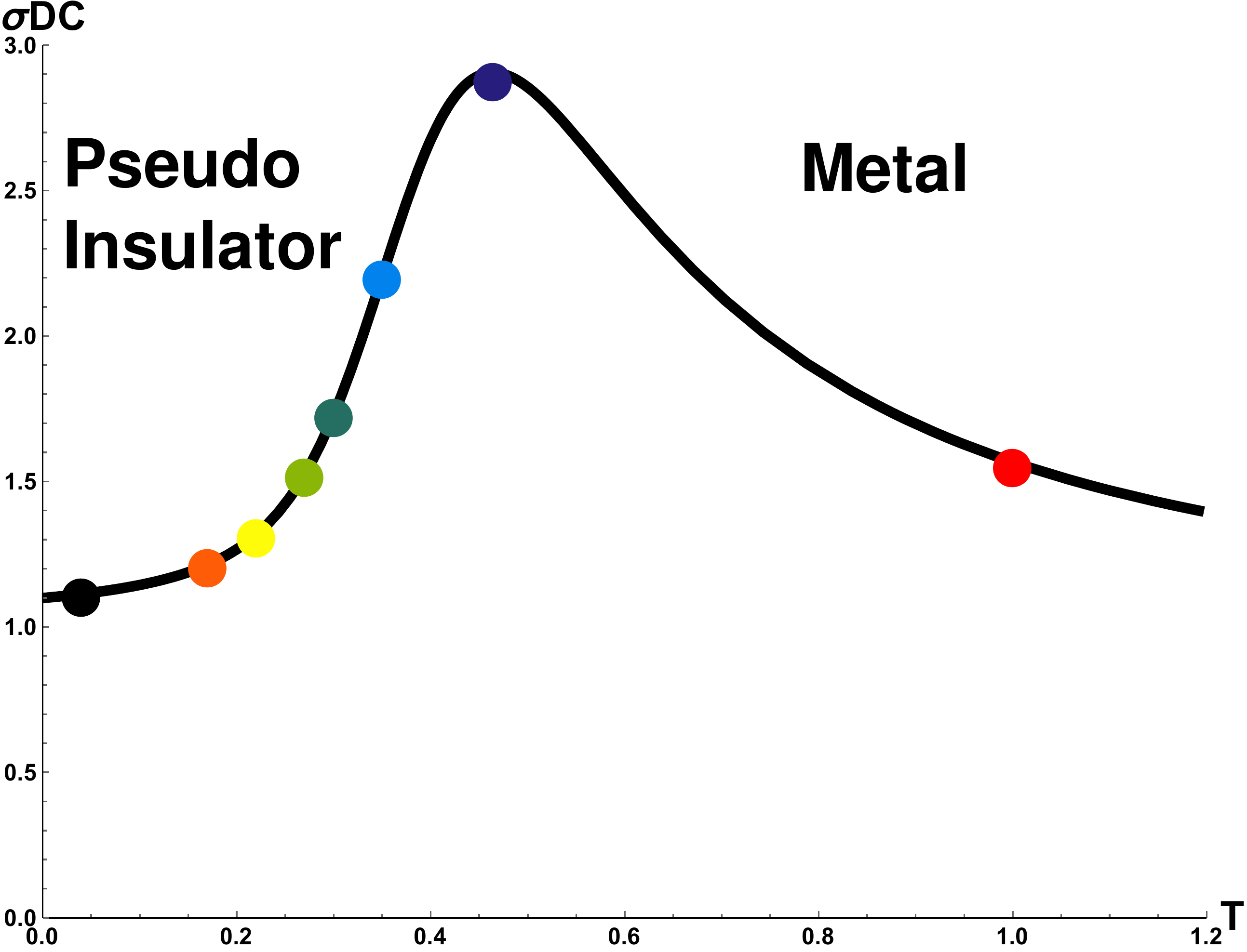}
\caption{\textbf{Left:} Optical conductivity in the normal phase for (\ref{model2}) ($\alpha=\sqrt{2}$, $m^2=0.025$, $\rho=1$) with temperature running from $T = 0.04$ (black line) to $T = 1$ (red line). \textbf{Right:} DC conductivity as a function of temperature for the same model and parameters. }
\label{NormalFeat}
\end{figure}

As suggested in \cite{Baggioli:2014roa}, for models with a specific
choice of the Lagrangian $V(X)$, the solution exhibits various interesting properties.\\
Using the membrane paradigm the DC part ($\omega=0$) of the optical conductivity can be computed analytically \cite{Blake:2013bqa} and for a generic Lagrangian $V(X)$ it is given by \cite{Baggioli:2014roa}:
\begin{equation}
\sigma_{DC}=\frac{1}{e^2}\left(1+\frac{\rho^2\,u_h^2}{2\,m^2\,\alpha^2\,\dot V(u_h^2\,\alpha^2)}\right)\,.
\label{DCformula}
\end{equation}
The DC conductivity consists of two parts:
\begin{equation}
\sigma_{DC}\,=\,\sigma_{pair}+\sigma_{dissipation}\,,
\end{equation}
which is a generic holographic feature
The first one $\sigma_{pair}$ is due to pair creation in the background, and it is present even at zero charge density \cite{Karch:2007pd}. It corresponds exactly to the probe limit result.
It is temperature independent, and therefore is always present (unless we introduce a dilaton field) as an offset in the value of $\sigma_{DC}$,
leading to $\sigma_{DC}(T=0)\neq 0$.
The second term
$\sigma_{dissipation}$ is really the one dealing with dissipative mechanism, and it can be thought as the strongly coupled analogue of the Drude formula for the conductivity. In the limit of zero translational symmetry breaking parameter $m$,
this second term gives rise to the infinite DC conductivity, typical for backgrounds preserving translational symmetry, such as the AdS Reissner-Nordstrom black brane case.
Due to the freedom of choice of the Lagrangian $V(X)$ this solution can be either a metal or a pseudo-insulator and can provide a transition between the two phases (see figure \ref{NormalFeat}).
The pseudo-insulator phase
is characterized by the conductivity, declining at smaller temperatures, $d\sigma/dT>0$,
for $T<T_0$,
but reaching a non-vanishing value at $T=0$ (which is the reason why we are not calling it an insulating phase)\footnote{
One easy way to enable $\sigma_{DC}(T=0)=0$ is
adding a dilaton field to the action \cite{Gouteraux:2014hca}, which allows to
get a ``real" insulating state. See also \cite{BPprogress} for an alternative approach.}.
The transition between the two phases is provided by the existence of  a maximum in the DC conductivity as a function of temperature (see figure \ref{NormalFeat}), at $T=T_0$, which gives a clear separation between two different regimes:
\begin{align}
& \frac{d\sigma}{dT}<0\,,\quad T>T_0 \,\,\rightarrow\,\,\text{metal}\label{metalder}\\
& \frac{d\sigma}{dT}>0\,,\quad T<T_0 \,\,\rightarrow\,\,\text{pseudo-insulator}\label{insulatorder}
\end{align}
The temperature $T_0$ at which the metal-insulator transition happens can be obtained analytically, solving the following equation:
\begin{equation}
\frac{d\sigma_{DC}}{du_h}=0\quad\Rightarrow\quad Y \ddot V(Y)\,=\,\dot V(Y)\,,\,\,\,\,\,\,\,\,\,Y\,=\,u_h^2\,\alpha^2\,.
\end{equation}

The metal-insulator transition in the behavior of the DC conductivity
is related to a non-trivial structure in the optical conductivity,
namely a weight transfer from a Drude peak into a localized new peak in the mid-frequency regime \color{black}(see figure
\ref{NormalFeat}). This feature corresponds to an emerging collective propagating excitation
of the charge carriers, whose nature is not completely clear yet.
The phase diagram of this normal phase is already rich and can give insights towards the interpretation about the various ingredients introduced into the model. In the case of the linear Lagrangian, which goes back to the original model \cite{Andrade:2014xca}, the parameters $m$ and $\alpha$ are combined into $m\,\alpha$,
which can be interpreted as the strength of translational symmetry breaking. From the dual field theory point of view this is thought to be related to some sort of homogeneously distributed density of impurities,
representing the disorder-strength in the material.\\

\begin{figure}
\centering
\includegraphics[width=.4\textwidth]{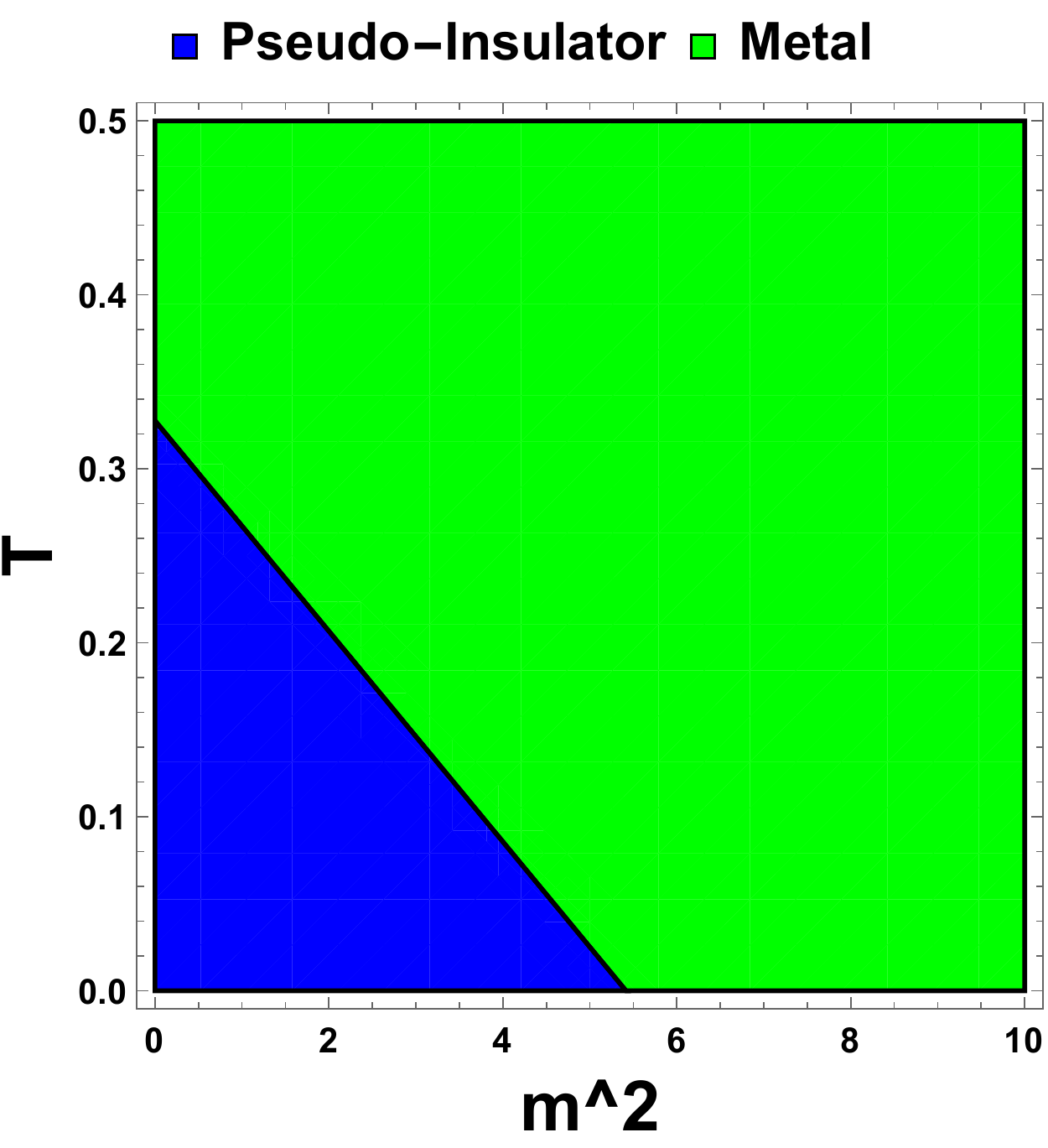}
\includegraphics[width=.4\textwidth]{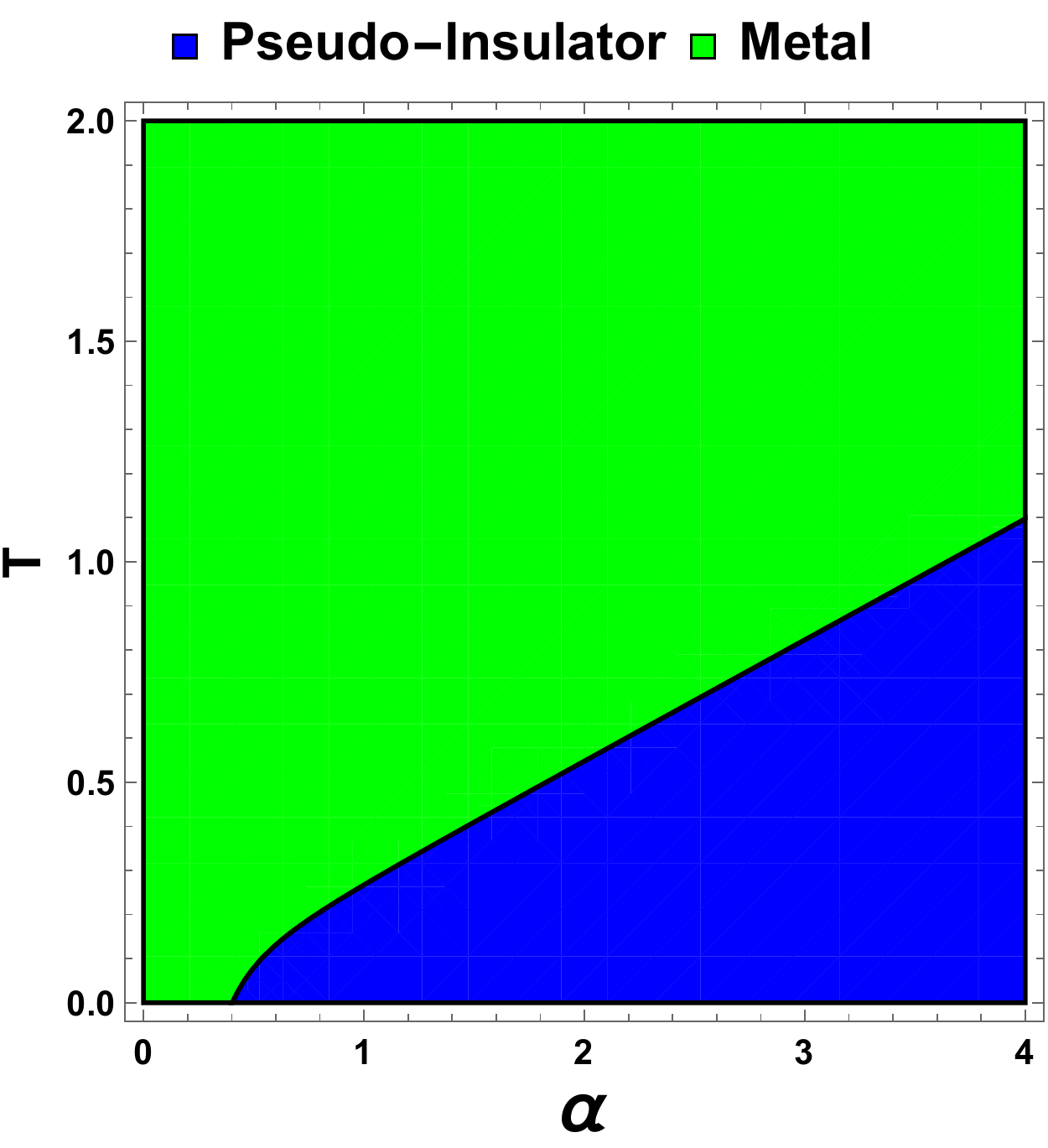}
\caption{Region plots for the model (\ref{model2}) in the normal phase. We choose
units where the density is $\rho=1$. Here we have fixed $\alpha=1$ (left plot) and $m=1$ (right plot).
The blue region is pseudo-insulating, $d\sigma_{DC}/dT>0$,
the green region is metallic, $d\sigma_{DC}/dT<0$.}
\label{NormalPhase}
\end{figure}

In the case of a more general $V(X)$, the $m$ parameter keeps this kind of interpretation while the $\alpha$ one represents the strength of interactions of the neutral scalar sector. This reasoning is confirmed by the study of the phase diagrams of the system (figure \ref{NormalPhase}) which makes evident the difference between the two parameters. Indeed, while the $m$ parameter, which we are going to interpret as the disorder-strength of our High-Tc superconductor, enhances the metallic phase, the $\alpha$ one clearly reduces the mobility of the electronic sector driving the system towards the pseudo-insulating phase.\color{black}

\section{Superconducting instability}\label{section3}

In this section we will describe the instability conditions for the normal phase towards the development
of a non-trivial profile of the charged scalar field. This allows one to determine a line of the second
order superconducting phase transition, $T_c(\alpha)$ (or $T_c(m)$
for the model (\ref{model2})), in the boundary field theory, with broken
translational symmetry.

We start by considering the system at zero temperature,
which we are able to study analytically. 
Then we proceed to studying the normal phase at a finite temperature.
Upon lowering the temperature, at a certain critical value $T=T_c$,
the normal phase becomes unstable. This is the point of
a superconducting
phase transition.
We construct numerically $T_c$ as a function of the parameters $\Delta$, $q$, $\alpha$ (or $m$),
for the models with various $V(X)$.

\subsection{Zero-temperature instability}

\begin{figure}
\centering
\includegraphics[width=.55\textwidth]{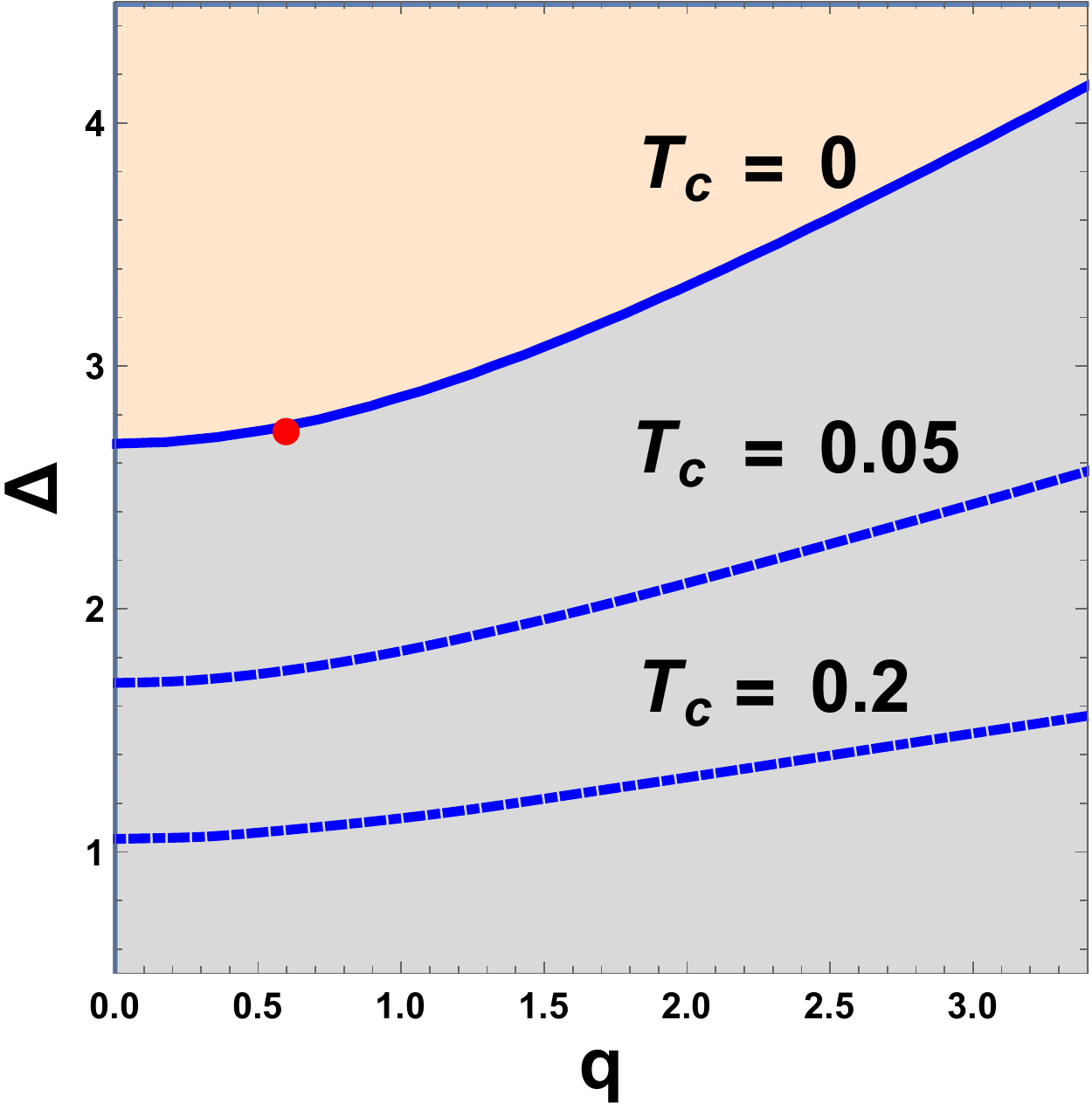}
\caption{Region and contour plots for the model (\ref{model1}) with linear
potential for the neutral scalars (dimensional quantities are measured in units of the charge
density $\rho$). We choose $\alpha=2$.
The region of $\Delta$, $q$, satisfying the IR instability condition (\ref{domez})
is shaded in grey. The red dot is centered around
$(q_d,\Delta_d)=(0.6,\,2.74)$. These tuned $(q,\Delta)$
confine superconducting phase of the model (\ref{model1})
into a dome region, as we discuss in Section \ref{section6}. Notice the proximity
of the red dot to the boundary of the IR instability region, resulting 
in $T_c(q_d,\Delta_d)$ being very small.}
\label{TcConsplot1}
\end{figure}

\begin{figure}
\centering
\includegraphics[width=.55\textwidth]{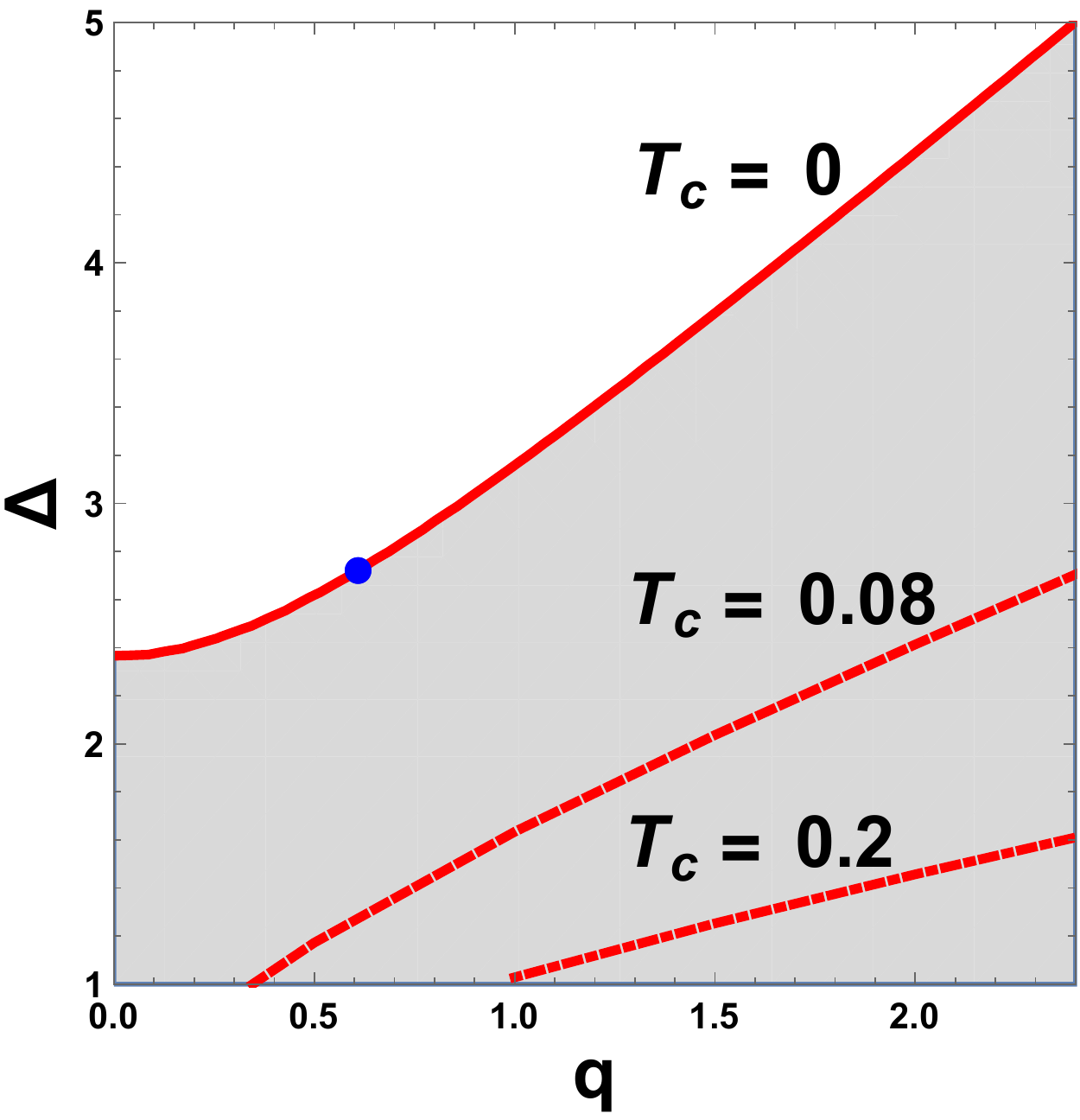}
\caption{Region and contour plots for the model (\ref{model2}) with non-linear
potential for the neutral scalars with $\alpha=0.25$, $m=4$ (dimensional quantities are measured in units of the charge
density $\rho$).
The region of $\Delta$, $q$, satisfying the IR instability condition (\ref{domez})
is shaded. The blue dot on the plot has coordinates
$(q_d,\,\Delta _d)=(0.6,\, 2.74)$. These tuned $(q,\,\Delta)$
confine superconducting phase of the model (\ref{model2})
into a dome region, as we discuss in Section \ref{section6}. Notice the proximity
of the blue dot to the boundary of the IR instability region, resulting 
in $T_c(q_d,\Delta_d)$ being very small.}
\label{TcConsplot2}
\end{figure}

In the case of $T=0$ the normal phase geometry interpolates
between the $AdS_4$ in the ultra-violet and the $AdS_2\times \mathbb{R}^2$
in the infra-red. We can apply the known analytical calculation to study the stability
of the normal phase towards formation of a non-trivial profile of the scalar $\psi$ \cite{Denef:2009tp}.

Due to eq. (\ref{scfeq}), the effective mass $M_{eff}$ of the scalar $\psi$ is given by:
\beq
M_{eff}^2\,L^2 =\,M^2\,L^2+\kappa\,H\,L^2+q^2\,A_t^2\, g^{tt}\,L^2\,.\label{psiHmass}
\eeq
Notice that at the boundary the mass of the scalar is just $M^2$ but at the horizon it gets an additional contribution.
This is because near the horizon we have:
\beq
g^{tt}=-\frac{2u_h^2}{L^2\,f''(u_h)(u_h-u)^2}\,,
\eeq
at zero temperature.
Due to (\ref{eomsnormalt}), and the zero temperature $T=0$ condition, with the temperature given by (\ref{normalT}), we obtain:
\begin{align}
f''(u_h)
=\frac{2\left(6+L^2m^2\left(-2V(u_h^2\alpha^2)+u_h^2\alpha^2\dot V(u_h^2\alpha^2)\right)\right)}{u_h^2}\,. \label{fpp}
\end{align}
The normal phase is unstable towards formation of the scalar hair, if $M_{eff}$ violates 
the BF stability bound in the $AdS_2$, namely:
\beq
M_{eff}^2\,L^2_{2}< -\frac{1}{4}\,.\label{IRinst}
\eeq
In (\ref{IRinst}) we have denoted the $AdS_2$ radius as $L_2$\footnote{In the usual RN case we have $f''(u_h)\,=\,\frac{12}{L^2u_h^2}$, and we find the usual $L_2^2\,=\,\frac{L^2}{6}$ in $d=3$.} :
\begin{equation}
L_2^2\,=\,\frac{2\,L^2}{f''(u_h)\,u_h^2}\label{L2}
\end{equation}
Combining (\ref{psiHmass}), (\ref{fpp}), (\ref{L2}), the IR instability condition
(\ref{IRinst}) finally reads\footnote{This formula agrees with \cite{Kim:2015dna} in the case of $V(X)=\frac{X}{2\,m^2}$ and $\kappa=0$.} :
\begin{equation}
D<0\,,\label{domez}
\end{equation}
where we have defined the function $D$ as:
\beq
\label{Ddef}
D=\frac{1}{4}+\frac{L^2 \left(\kappa  H+M^2\right) \left(L^2 m^2 \left(\alpha ^2 u_h^2 \dot V\left(\alpha ^2 u_h^2\right)-2 V\left(\alpha ^2 u_h^2\right)\right)+6\right)-q^2 \rho ^2 u_h^4}{\left(L^2 m^2 \left(\alpha ^2 u_h^2 \dot V\left(\alpha ^2 u_h^2\right)-2 V\left(\alpha ^2 u_h^2\right)\right)+6\right)^2}
\eeq
For the practical calculations we will solve the equation $T=0$, see (\ref{normalT}), for the value
of $u_h$, giving  the position of the horizon of the extremal black brane,
\beq
\label{ztempc}
-12+u_h^4\,\rho^2+4L^2\,m^2\,V(u_h^2\,\alpha^2)=0\,.
\eeq
We will measure all the dimensional quantities in units of $\rho$;
both for zero temperature and finite-temperature instability analyses the $\rho$
can be scaled out.

In figure \ref{TcConsplot1} we plot the IR instability region on the $(\Delta,q)$ plane,
for the model~1, (\ref{model1}), with $\alpha=2$, as well as a few contour lines of the constant
critical temperature. In figure \ref{TcConsplot2} we plot the IR instability region and several $T_c={\rm const}$
curves on the $(\Delta,q)$ plane,
for the model~2, (\ref{model2}), with $\alpha=0.25$, $m=4$.
Analogous plot for ordinary holographic superconductor can be found in \cite{Denef:2009tp}.
Plot in the case of the linear $V(X)$ model first appeared in \cite{Kim:2015dna}.

\subsection{Finite-temperature instability}\label{subfininst}

\begin{figure}
\centering
\includegraphics[width=.35\textwidth]{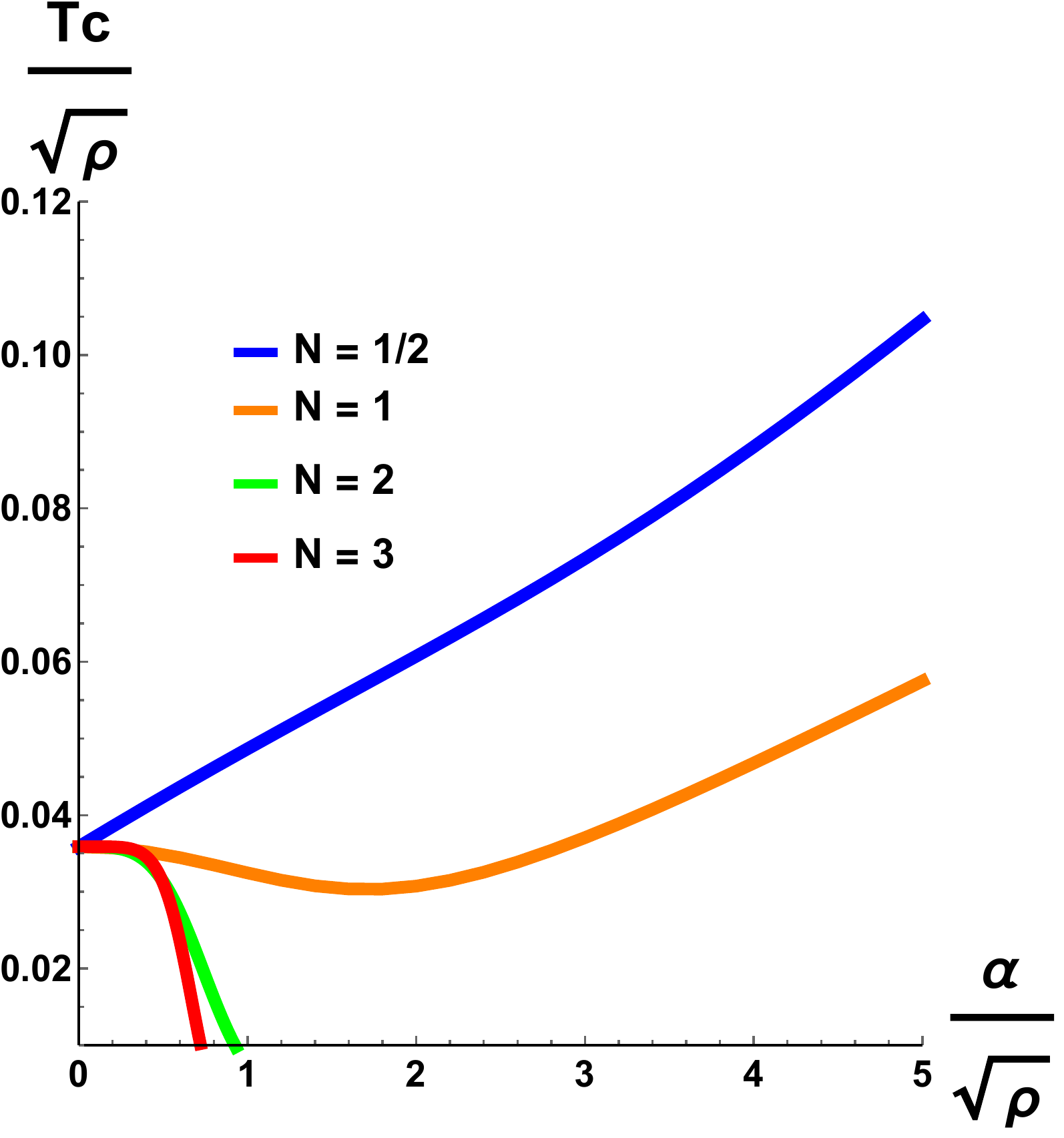}\qquad\qquad
\includegraphics[width=.35\textwidth]{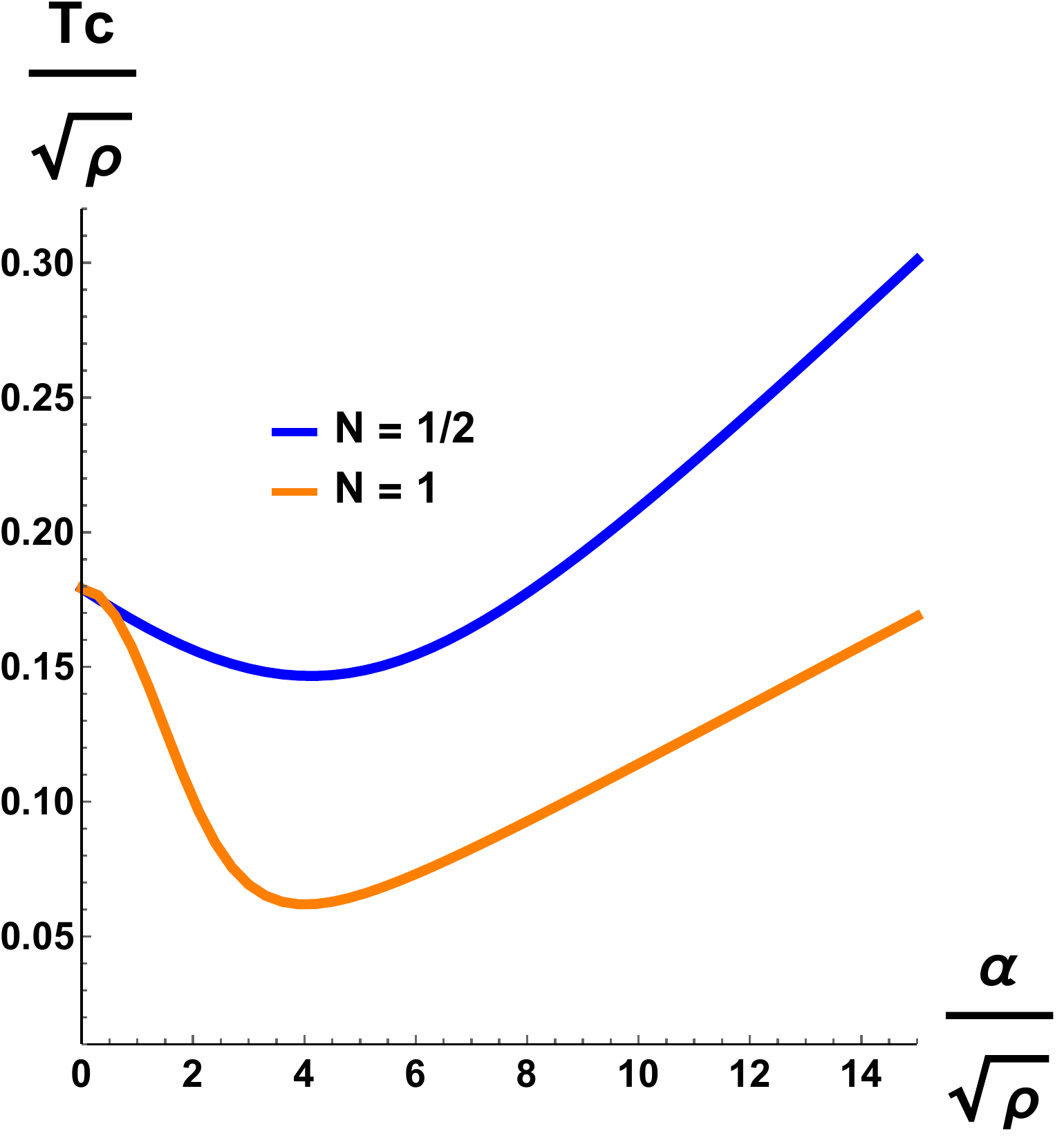}
\caption{Critical temperature as a function of $\alpha$ for the model (\ref{model3}). \textbf{Left:} 
$q=1$, $\Delta=2$. \textbf{Right:} $q=3$, $\Delta=2$.}
\label{figTc1}
\end{figure}
\begin{figure}
\centering
\includegraphics[width=.35\textwidth]{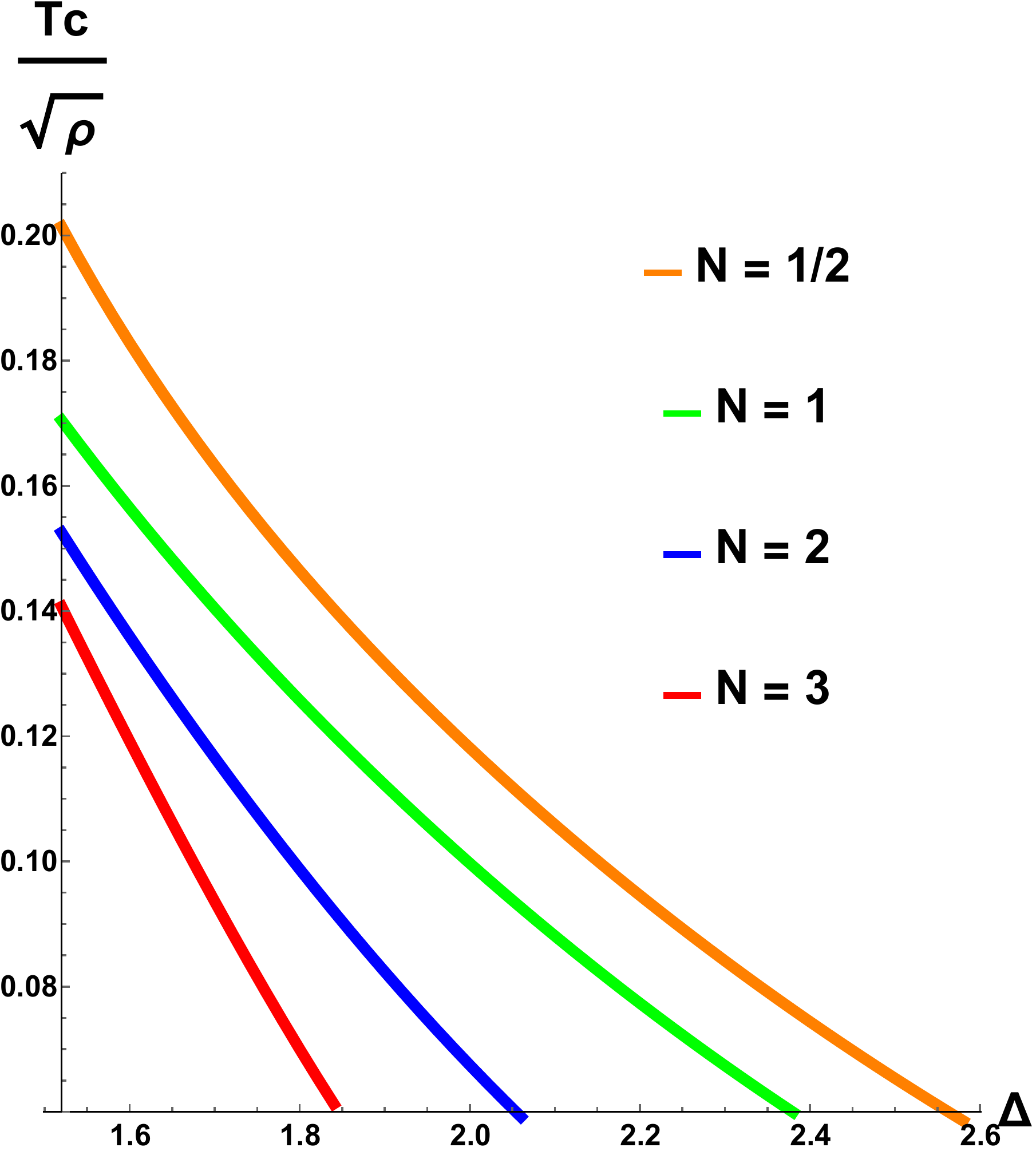}\qquad\qquad
\includegraphics[width=.35\textwidth]{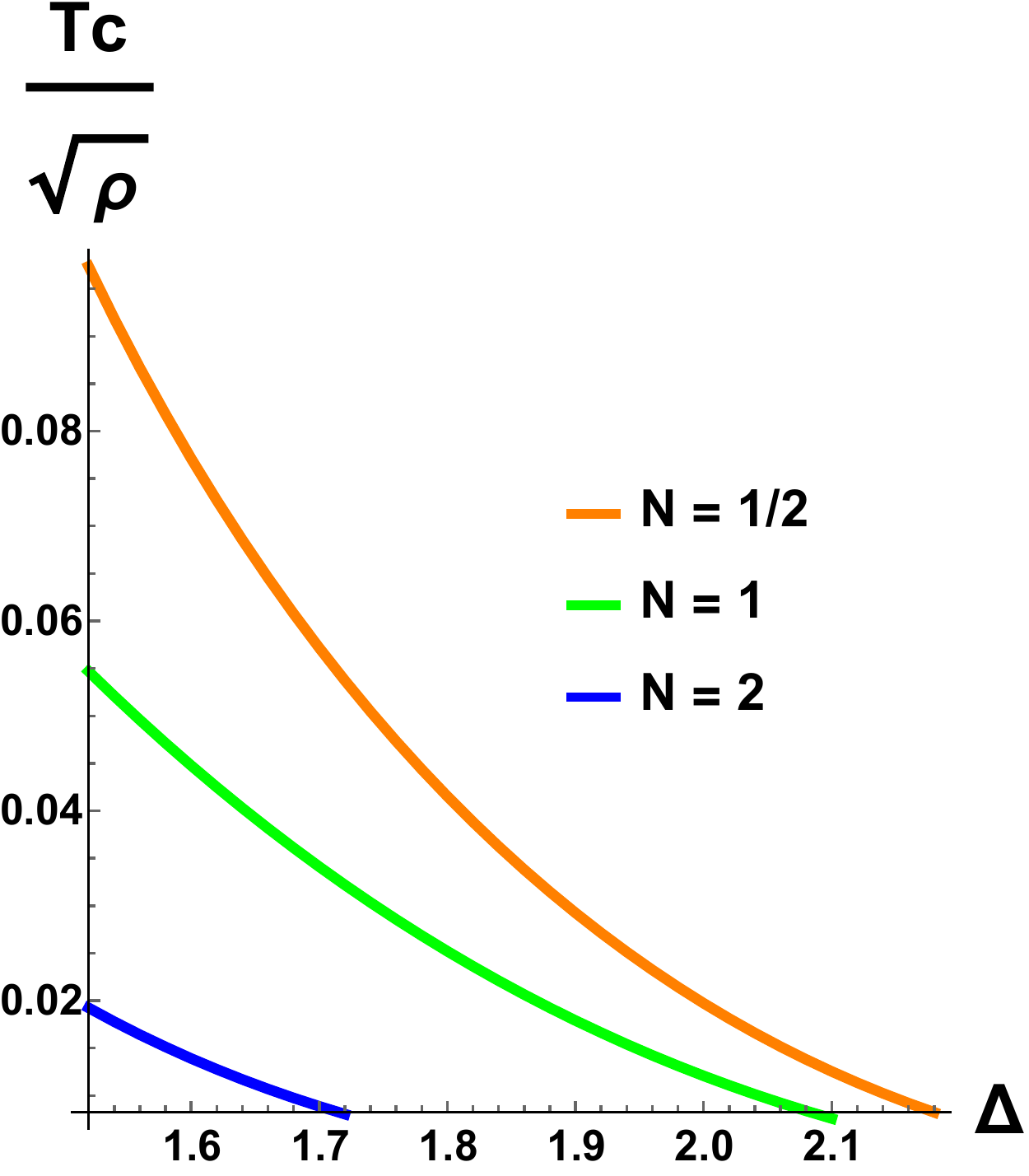}
\caption{Critical temperature as a function of $\Delta$ for the model (\ref{model3}). \textbf{Left } $(\alpha=1,\,q=2)$. \textbf{Right:} $(\alpha=1,\,q=0.6)$.}
\label{figTc2}
\end{figure}\begin{figure}
\centering
\includegraphics[width=.45\textwidth]{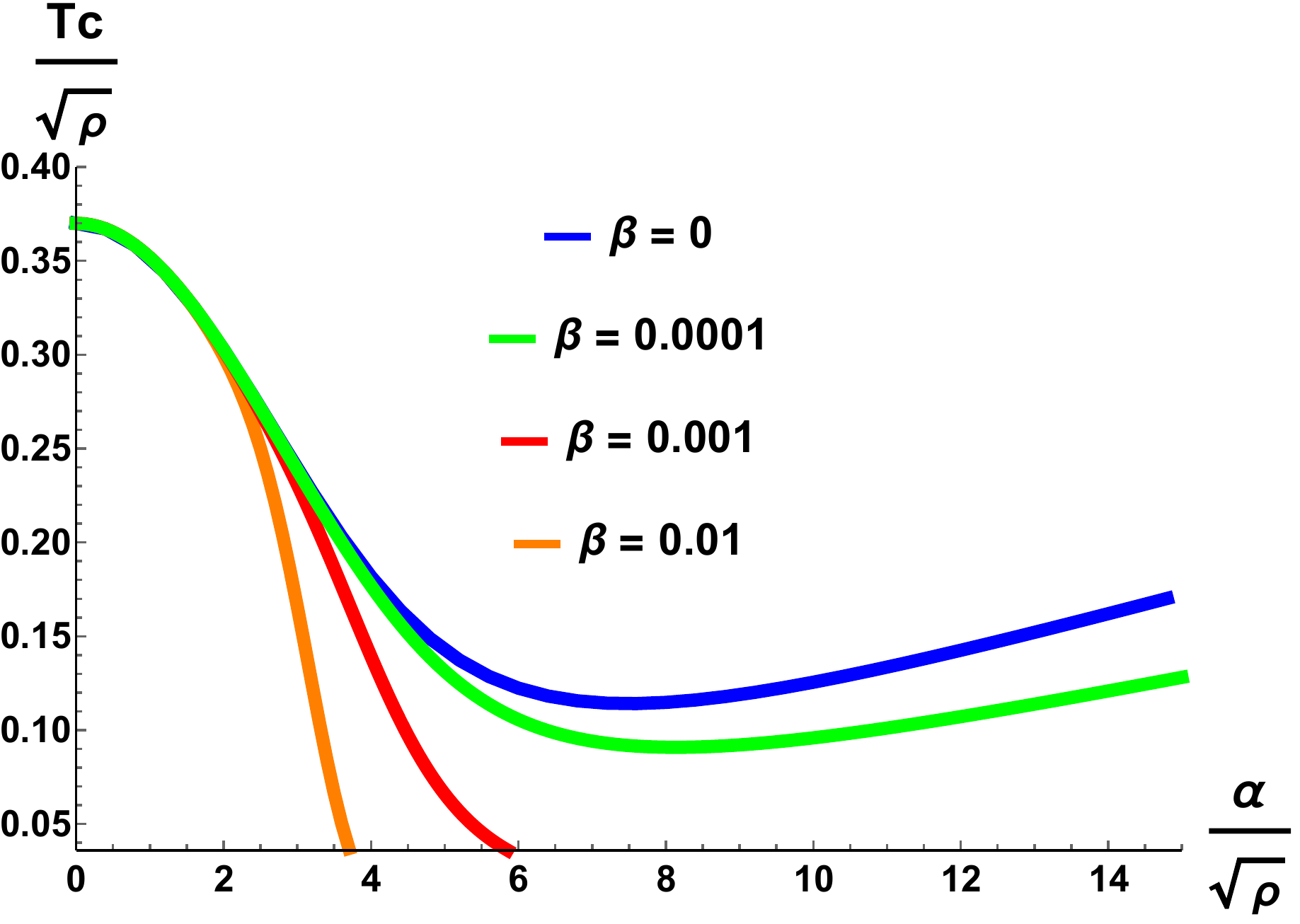}
\caption{Critical temperature as a function of $\alpha$ for the potential $V(X)= X/2m^2+\beta\,X^5/2m^2$ for different choices $\beta$. All the curves have a runaway behavior at $\alpha\rightarrow\infty$,
and only the shape depends on the value of $\beta$.}
\label{figTc3}
\end{figure}

Consider the system at large temperature in a normal phase,
which exists in a superconducting phase at low temperatures.
Therefore as we decrease the temperature, at certain critical value $T_c$ the superconducting phase transition occurs. If $T_c$ is non-vanishing, then for $T<T_c$ the system is in a superconducting phase,
with a non-trivial scalar condensate $\psi(u)$. 

Recall that near the boundary the scalar field with mass $M$:
\beq
M=\frac{1}{L}\sqrt{\Delta(\Delta-3)}\,,
\eeq
behaves asymptotically as:
\beq
\psi(u)=\frac{\psi_1}{L^{3-\Delta}}u^{3-\Delta}+\frac{\psi_2}{L^\Delta}u^\Delta\,,\label{nbexp}
\eeq
where $\psi_1$ is the leading term, identified as the source in the standard quantization.

To find the value of $T_c$
we can look for an instability of the normal phase towards formation of the scalar field profile \cite{Denef:2009tp,Amado:2009ts}.
Near the second order phase transition point $T=T_c$
the value of $\psi$ is small, and therefore one can neglect its backreaction on
the geometry. 
The SC instability can be detected by looking at the motion of the QNMs of $\psi$ in the complex plane. To be more specific, it corresponds to a QNM going to the upper half of
a complex plane.
Exactly at critical temperature we have a static mode at the origin of the complex plane, $\omega=0$, and the source at the boundary vanishes, $\psi_1=0$.
In the next section we will solve numerically
the equations (\ref{sEinst})-(\ref{scfeq}) for the whole
background, and confirm this explicitly.

The scalar field is described by eq. (\ref{scfeq}), which in the normal phase becomes:
\beq
\psi''+\left(-\frac{2}{u}+\frac{f'}{f}\right)\psi'+\left(\frac{q^2\rho^2}{f^2}-\frac{M^2 L^2}{u^2 f}-\frac{\kappa H\,L^2 }{u^2 f}\right)\psi\,=0\,,\label{scfeqn}
\eeq
where $f(u)$ is given by (\ref{eomsnormalt}).
To determine the critical temperature $T_c$ we need to find the {\it highest} temperature,
at which there exists a solution to eq. (\ref{scfeqn}), satisfying the $\psi_1=0$ condition.
In this case for $T<T_c$ the system is in a superconducting state, with a non-vanishing condensate $\psi_2$.\\

We are interested in the phases of the models
(\ref{model1})-(\ref{model3}) on the temperature-disorder strength plane.
\color{black}
In figure \ref{figTc1} we plot $T_c(\alpha)$ for the model~1, (\ref{model1}),
and the model 3, (\ref{model3}), with $N=1/2,2,3$, for different values of the charge $q$. It is clear that
when the power $N$ in the potential $V$ is higher, the critical temperature for the SC phase
transition is smaller. One interesting behavior, which still lacks an interpretation, is the non-monotonic behavior of $T_c$ as a function of $\alpha$, which was already observed in the original model \cite{Andrade:2014xca,Kim:2015dna} and still persists in more generic setups.

In figure \ref{figTc2} we plot $T_c(\Delta)$ for $q=0.6$ and $\alpha=1$
for the model~1, (\ref{model1}),
and the model 3, (\ref{model3}), with $N=1/2,2,3$. 
The $T_c(\Delta)$ curves explicitly show that the critical
temperature quickly declines as $\Delta$
approaches the border of the IR instability region.
It is further underlined how higher powers/non-linearities in the potential lead to deeper supression of the critical temperature.

We also plot
$T_c(\alpha)$ for the generalized model 2, (\ref{model2}), in figure \ref{figTc3} for various amounts
of non-linearity $\beta\,X^5$, showing again the same behavior of suppression of the
superconducting phase at larger $\beta$.

\section{Broken phase and phase diagram}\label{section4}

In this section we study superconducting phase and construct the phase diagram
on
the $(m,T)$ plane of the model
(\ref{model2}).
We will confirm existence of the second order phase
transition between normal and superconducting phases
by solving four equations (\ref{sEinst})-(\ref{scfeq}) for the fully backreacted background.
Knowing the near-boundary asymptotic behavior of this solution, one can determine the grand potential
of the superconducting phase, and compare it with the grand potential of the normal phase
to corroborate the phase transition at $T=T_c$.

\begin{figure}
\begin{center}
\includegraphics[width=.45\textwidth]{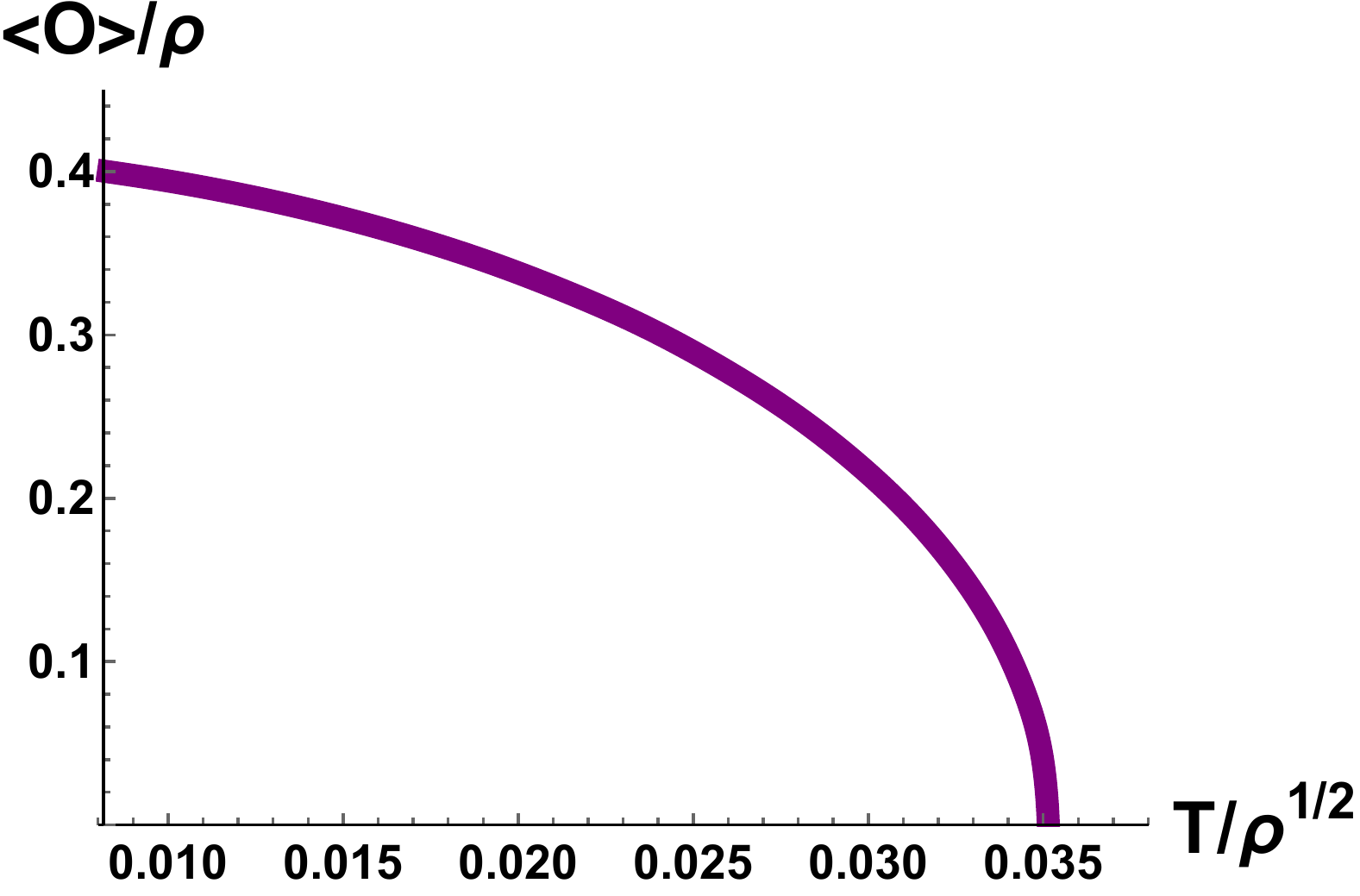}
\includegraphics[width=.45\textwidth]{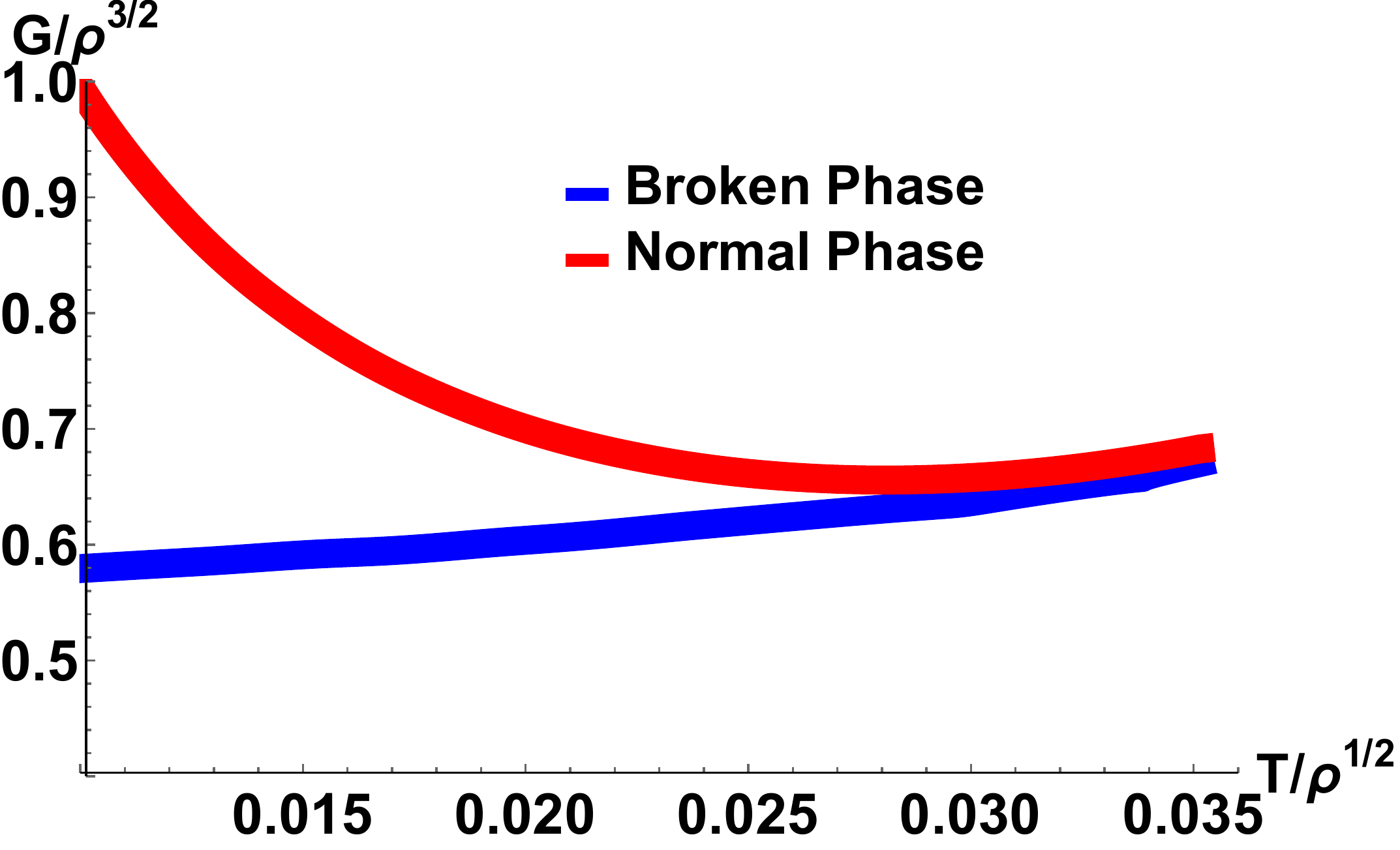}
\end{center}
\caption{Condensate for $\Delta=2$, $q=1$, $\alpha\,u_h=0.5$, $m\,L=1$ model
with $V=z+z^5$, and the corresponding Grand Potential for the two phases.}
\label{fig:ConductivityNeutralk=0}
\end{figure}

Running the numerical procedure described in details in appendix \ref{appendix2}, we were able to construct the condensate
$\psi_2/\rho^{\Delta/2}$ as a function of temperature $T/\rho^{1/2}$.
In figure \ref{fig:ConductivityNeutralk=0} we provide the plot of the condensate, for the model (\ref{model2}) with $\Delta=2$,
$q=1$, $\alpha\,u_h=0.5$ ($\alpha$ in units of entropy density), $m\,L=1$.
There we also plot
the grand potential
for the broken and normal phases which confirms the superconducting transition at $T=T_c$.

The holographic prescription for the calculation of the grand potential is:
\begin{equation}
\Omega\,=\,-T\,\log Z\,=\,T \mathcal{S}_{E}\,,
\end{equation}
where $\mathcal{S}_{E}$ is a Euclidean on-shell action of the bulk theory.

After some computations showed in details in appendix \ref{appendix2} we obtain:
\beq
\mathcal{S}_{E}=\int d^3x\,\left(16\pi\,S\,T+2L^2\gamma_2+L^2\mu\rho\right)\,,
\label{onsht}
\eeq
where we have also used the area-law expression for the entropy:
\beq
S=\frac{L^2}{4u_H^2}\,,
\eeq
The grand potential is finally given by:
\beq
\Omega=P\,{\cal V}=-\frac{1}{16\pi}\mathcal{S}_{E}\,,
\eeq
where ${\cal V}$ is a volume of spatial region.
In conclusion we obtain (denoting $\hat\rho=\rho L^2$) the expected thermodynamic relation:
\beq
P=\epsilon-TS-\mu\hat \rho\,.
\eeq
where the energy density of the system is given by:
\beq
\epsilon=-2\gamma_2 L^2\,.
\eeq

\begin{figure}
\centering
\includegraphics[width=.44\textwidth]{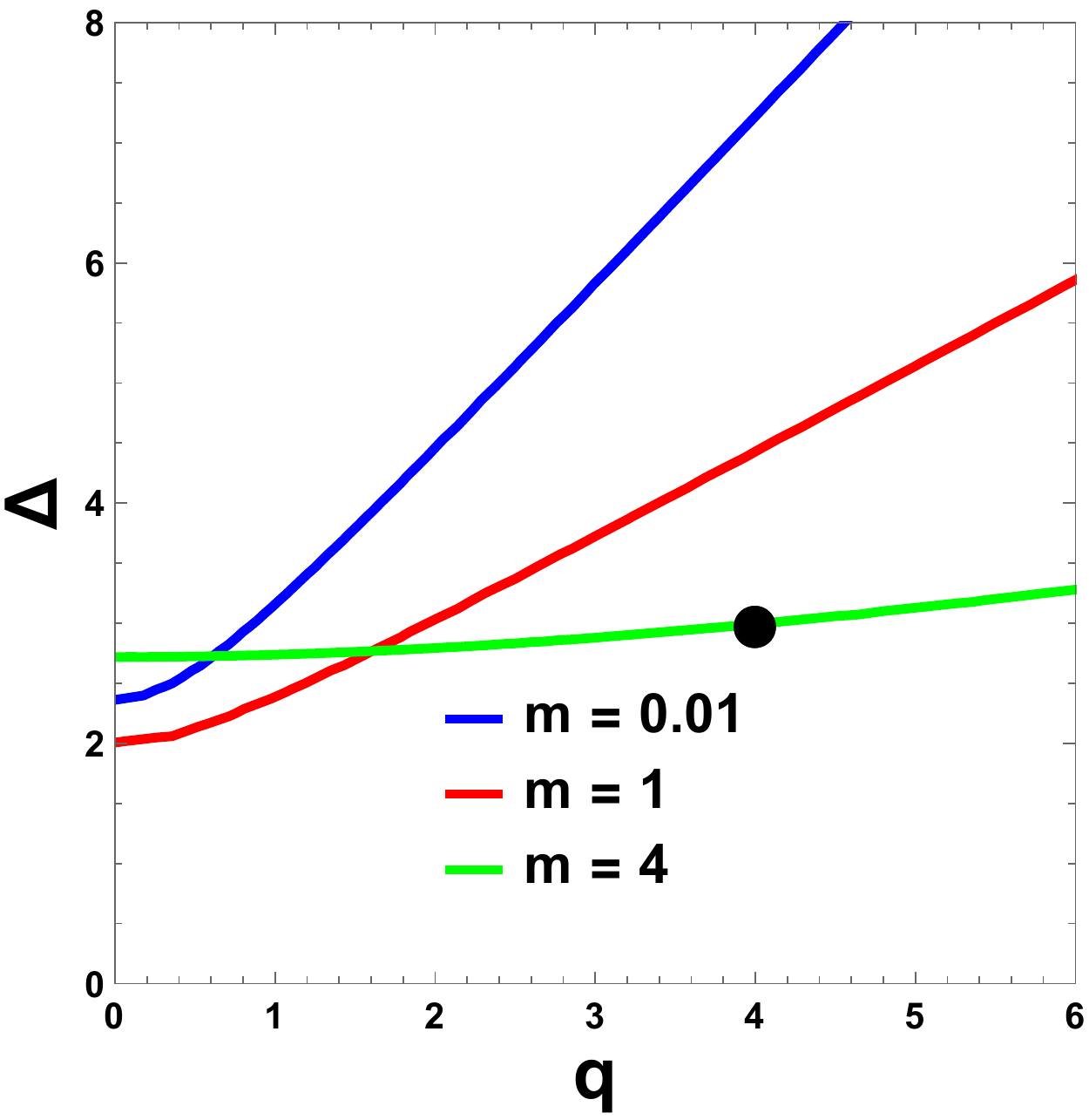}
\includegraphics[width=.46\textwidth]{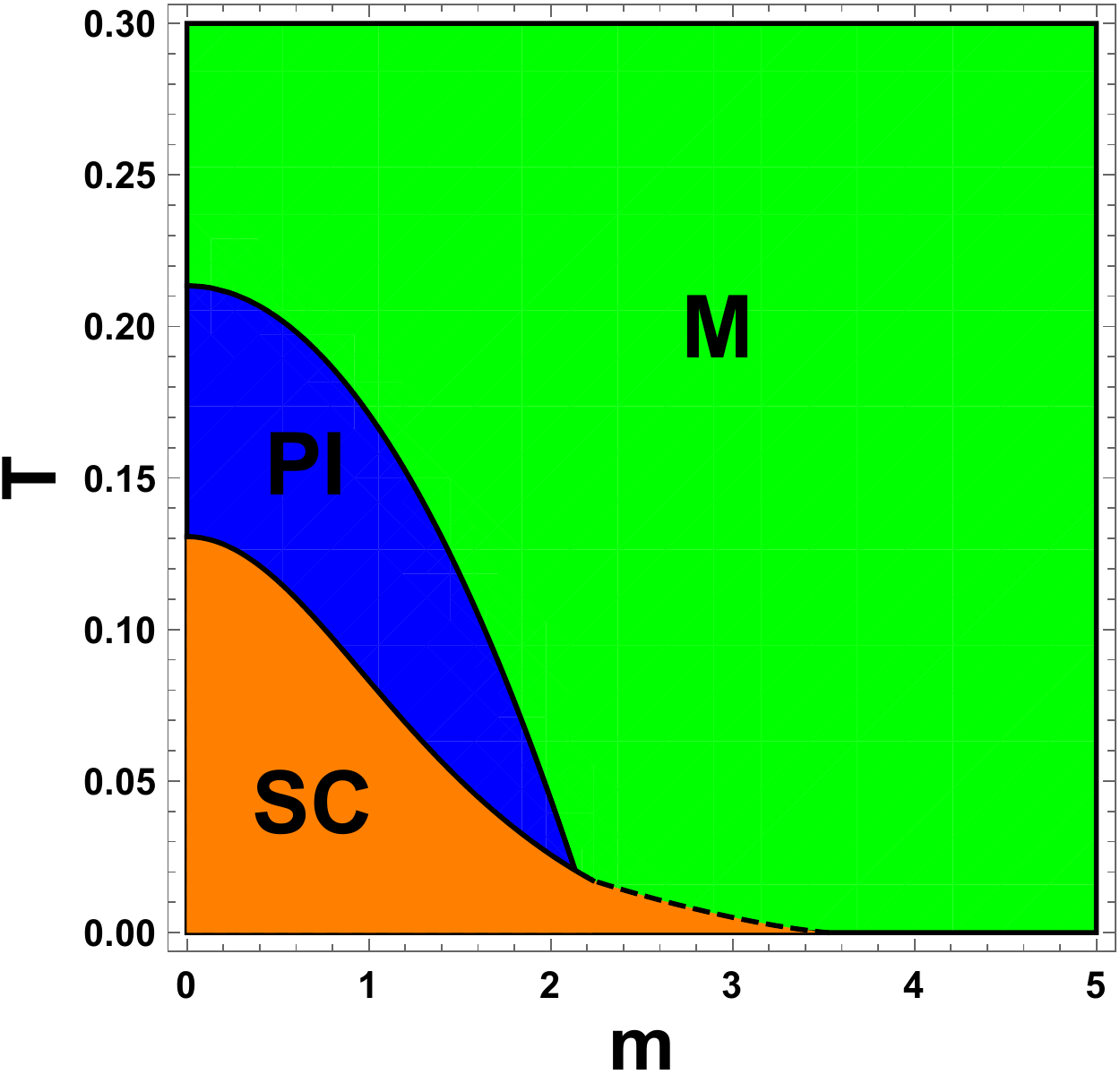}
\caption{{\bf Left:} The $T=0$ contour plots of the IR instability lines,
for the model (\ref{model2}) with $\alpha=0.7$ and various choices of $m$.
{\bf Right:} Phase diagram for the model (\ref{model2}) at the point $q=4$, $\Delta=3$ with $\alpha=0.7$ .
We have chosen units $\rho=1$.}
\label{PhaseDPlot}
\end{figure}

We now have enough information to construct the full phase diagram
of the non-linear model (\ref{model2}).
In figure \ref{PhaseDPlot} we plot the phase diagram
of the model (\ref{model2}) with $\Delta=3$, $q=4$, $\alpha=0.7$ (in units of $\rho=1$). We see that the superconducting region can be connected smoothly to both a metallic phase and a pseudo-insulating 
phase.

\section{Optical conductivity }\label{section5}

Our main aim in this section
is to see whether the non-trivial structure in the optical conductivity (see figure \ref{NormalFeat}),
pointed out for the model (\ref{model2}) in the normal phase \cite{Baggioli:2014roa},
persists to exist in the superconducting phase.

\subsection{Fluctuation equations}
In order to compute the optical conductivity, we study the fluctuations
on top of the charged black brane background with spatially-dependent neutral scalars, as follows:
\begin{align}
&\delta g_{tx}(t,u,y)\,=\,\int_{-\infty}^{+\infty}\frac{d\omega\,dk}{(2\pi)^2}\,e^{-i\,\omega\,t\,+\,i\,k\,y}\,\,\frac{h_{tx}(u)}{u^2}\notag\\
&\delta \phi_{x}(t,u,y)\,=\,\int_{-\infty}^{+\infty}\frac{d\omega\,dk}{(2\pi)^2}\,e^{-i\,\omega\,t\,+i\,k\,y}\,\,\xi(u)\label{fls}\\
&\delta A_{x}(t,u,y)\,=\,\int_{-\infty}^{+\infty}\frac{d\omega\,dk}{(2\pi)^2}e^{-i\,\omega\,t\,+\,i\,k\,y}\,a_x(u)\notag
\end{align}
We consider homogeneous perturbations defined by $k=0$ \color{black}, for which it is consistent
to put all the fluctuations, besides (\ref{fls}), to zero. In this section we also put $L=1$.
The equations for the perturbations read\footnote{In the case $V(X)=\frac{X}{2 m^2}$ these equations  reduce to the fluctuation
equations obtained in \cite{Kim:2015dna}.} :
\begin{align}
&a_x''+\left(\frac{f'}{f}-\frac{\chi'}{2}\right)a_x'+\left(\frac{e^{\chi}\omega^2}{f^2}-
\frac{2\,q^2\,\psi^2}{u^2\,f}\right)a_x+\frac{e^{\chi}A_t'}{f}h'_{tx}=0\\
&u^2\,a_x\,A_t'\,+\,h'_{tx}\,+\,\frac{2\,i\,e^{-\chi}\,m^2\,\alpha\,f\,\dot V(u^2\,\alpha^2)}{\omega}\xi'\,=\,0
\label{htxeq}\\
&\xi''\,+\,\left(-\frac{2}{u}+\frac{f'}{f}-\frac{\chi'}{2}+\frac{2\,u\,\alpha^2\,\ddot V(u^2\alpha^2)}{\dot V(u^2\alpha^2)}\right)\,\xi'\,+\,\frac{e^{\chi}\omega^2}{f^2}\xi\,-\,\frac{i\,e^{\chi}\,\alpha\,\omega}{f^2}h_{tx}\,=\,0
\end{align}

One can eliminate $h_{tx}'$ from the second equation (\ref{htxeq}) right away,
and substitute it into equations for $a_x$ and $\xi$ \cite{Andrade:2013gsa}. 
It is then convenient to perform the following redefinition:
\begin{equation}
\zeta(u)\,=\,\frac{f(u)}{i\,\omega\,\alpha\,u^2}\xi'(u)\label{zetaxi}
\end{equation}
and reduce the problem to a 2x2 system:
\begin{align}
&\left(e^{-\frac{\chi}{2}}\,f\,a_x'\right)'{+}e^{-\frac{\chi}{2}}\left({-}\frac{2q^2\psi^2}{u^2}{+}\frac{e^\chi (\omega^2
{-}u^2\,f\,A_t^{\prime 2})}{f}\right)\,a_x{+}2m^2\alpha^2u^2e^{-\frac{\chi}{2}}\dot VA_t'\zeta{=}0\label{flax}\\
&\left(\frac{u^2\,f\,e^{-\frac{\chi}{2}}}{\dot V}\left(e^{-\frac{\chi}{2}}\dot V\,\zeta\right)'\right)'
+u^2A_t'a_x+\left(\frac{\omega^2\,u^2}{f}-2m^2\alpha^2e^{-\chi} \dot V\zeta\right)\zeta=0\,,\label{flzeta}
\end{align}
which in the normal phase agrees with the equations, derived in \cite{Baggioli:2014roa}.

\subsection{Superconducting phase}
In order to extract the optical conductivity of the system we first derive the on-shell action for the fluctuations. We leave the technical steps for appendix \ref{appendix1} while here we just quote the result:
\beq
I_{tot}^f=\int d\omega\,\left(a_x^{(1)}(-\omega),\, Z^{(1)}(-\omega)\right)\,{\cal M}\,
\left({a_x^{(2)}(\omega\color{black})\atop Z^{(2)}(\omega\color{black})}\right)\,,
\eeq
where $\zeta= Z/u$ and we have defined the matrix $\cal M$ to be:
\beq
{\cal M}=\left({1\atop 0}\;
{0\atop
 \frac{2m^2\alpha^2V_1}{\sqrt{1-2\sqrt{2}+\frac{\sqrt{2}\omega^2}{m^2\alpha^2V_1}}}}\right)\,,
 \label{mmbulk}
\eeq
and expanded the fluctuations near the boundary $u=0$ as:
\begin{align}
a_x(u,\omega)&=a_x^{(1)}(\omega)+a_x^{(2)}(\omega)\,u\,,\\
Z(u,\omega)&=\frac{f(u)}{i\omega\alpha u}\,\xi'(u,\omega)\,,\\
Z(u,\omega)&=Z^{(1)}(\omega)+Z^{(2)}(\omega)\,u\,.
\end{align}

\begin{figure}
\centering
\includegraphics[width=.43\textwidth]{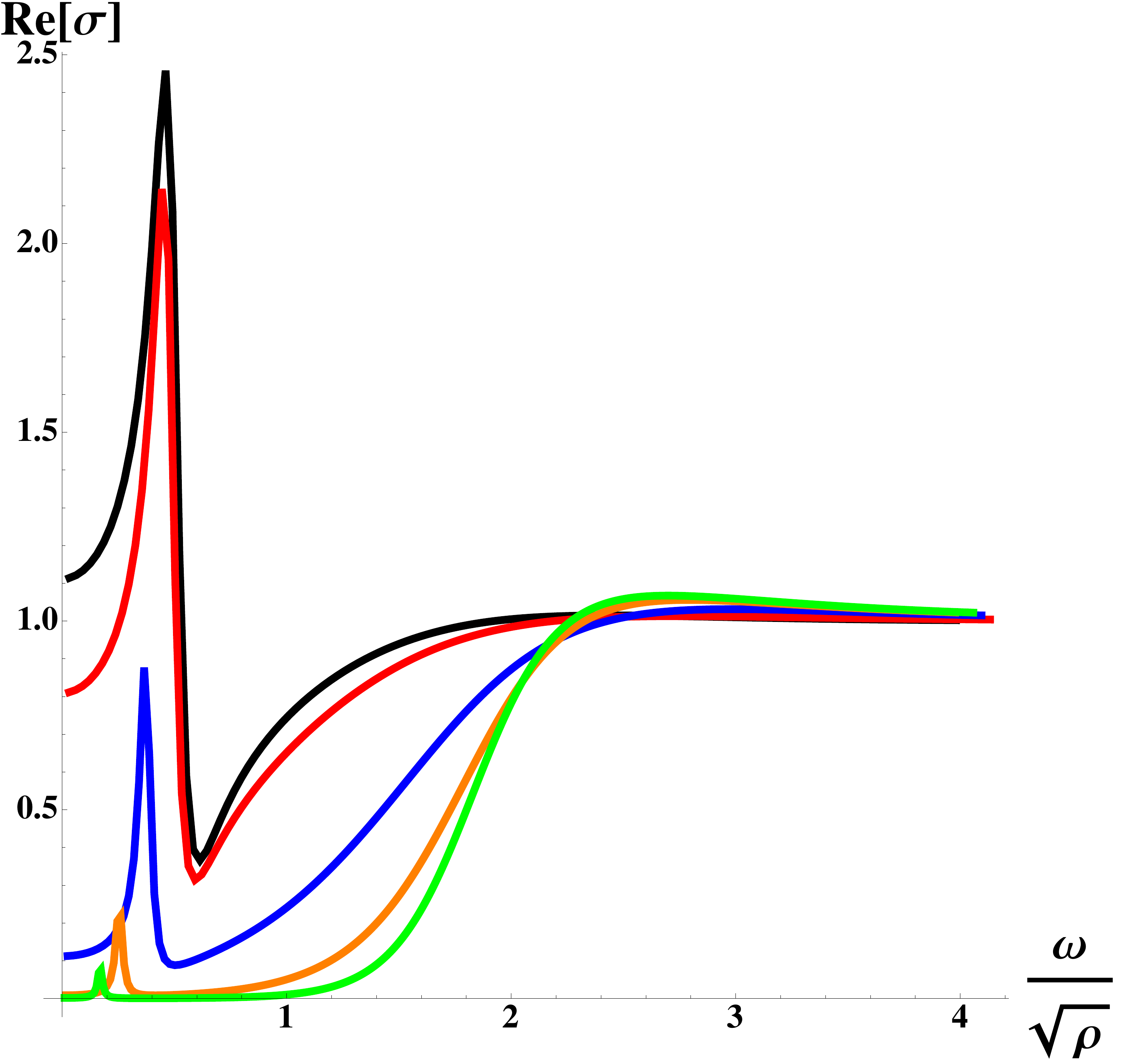}\qquad\qquad
\includegraphics[width=.43\textwidth]{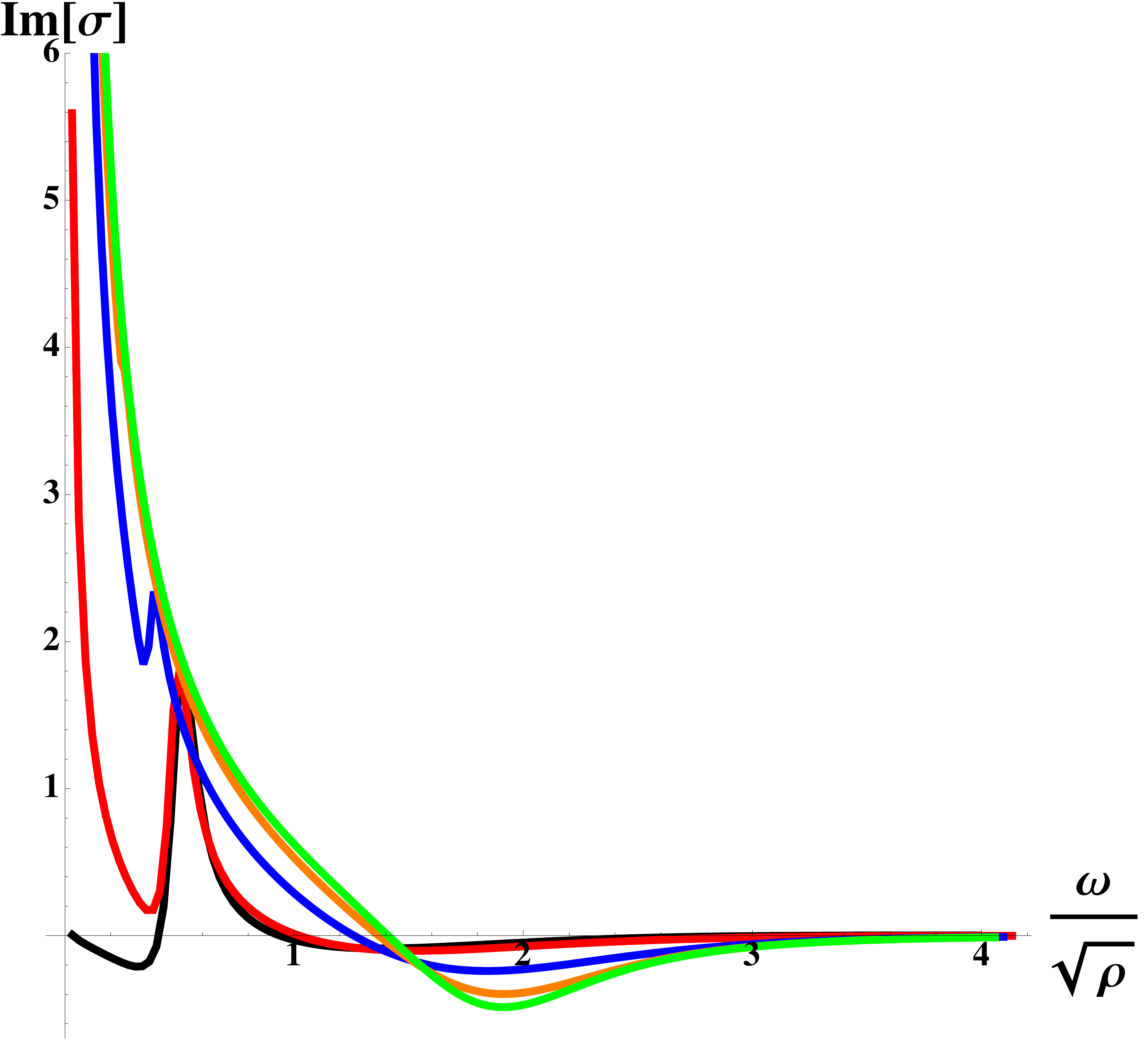}
\includegraphics[width=.43\textwidth]{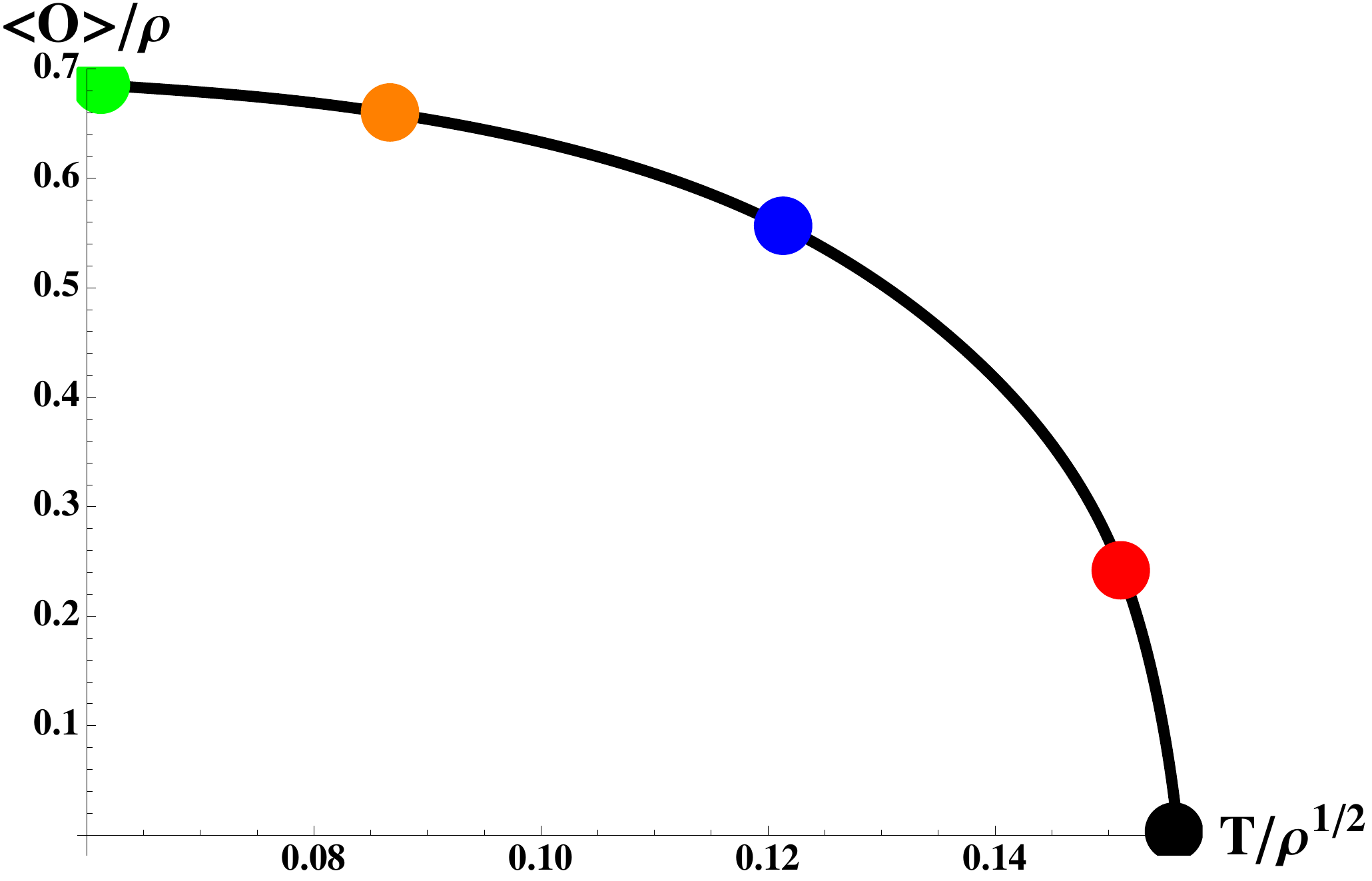}
\caption{The AC conductivity for the model (\ref{model2})
with $\alpha=\sqrt{2}$ (in units of $1/u_h$), $m^2L^2=0.025$, $q=4$, and $\Delta=2$.
Black line is at the temperature,
slightly below the corresponding critical temperature $T_c/\rho^{1/2}\simeq 0.16$, and
matches the result of the normal phase calculation at $T=T_c$.
Red, blue, orange and green lines are for $T/\rho^{1/2}=0.15, 0.12, 0.09, 0.06$, respectively. Notice that as we decrease the temperature, between blue and
orange line, the peak in the imaginary part of the AC conductivity disappears.
We call the corresponding critical temperature $T''/\rho^{1/2}\simeq 0.1$.
We also provide the condensate as a function of temperature and mark
the points where we calculated the AC conductivity.}
\label{ACplot}
\end{figure}

We solve two coupled fluctuation equations (\ref{flax}), (\ref{flzeta}) numerically,
for two independent sets of initial conditions which satisfy the infalling behavior near the horizon \cite{Kaminski:2009dh}\footnote{See also \cite{Goykhman:2012vy} for an example
of calculation of the correlation matrix in a different system of two coupled
fluctuations.}.
Due to linearity of the fluctuation equations (\ref{flax}), (\ref{flzeta}),
the precise choice of the two sets of initial conditions is not important,
and one can check that correlation matrix does not depend on it.
For example, let us choose:
\begin{align}
\left({a_x^{(1)}\atop Z^{(1)}}\right)&=\left({1\atop 1}\right)\,(u_h-u)^{\frac{i\omega}{f'(u_h)}e^{\chi_h/2}}\\
\left({a_x^{(2)}\atop Z^{(2)}}\right)&=\left({1\atop -1}\right)\,(u_h-u)^{\frac{i\omega}{f'(u_h)}e^{\chi_h/2}}\,.
\end{align}
Near the boundary the fields behave as:
\beq
\left({a_x^{(j)}\atop Z^{(j)}}\right)=\left({A_a^{(j)}\atop A_Z^{(j)}}\right)+\left({B_a^{(j)}\atop B_Z^{(j)}}\right)\,u\,,\qquad j=1,2\,.
\label{aZot}
\eeq
and we can assemble the matrices of leading and subleading coefficients:
\beq
{\cal A}=\left({A_a^{(1)}\atop A_Z^{(1)}}\;{A_a^{(2)}\atop A_Z^{(2)}}\right)\,,\quad
{\cal B}=\left({B_a^{(1)}\atop B_Z^{(1)}}\;{B_a^{(2)}\atop B_Z^{(2)}}\right)\,.\label{AABB}
\eeq
We collect the entries of the matrices (\ref{AABB}) by integrating the equations for the fluctuations
numerically and extracting the asymptotic behavior using (\ref{aZot}).
Knowing (\ref{AABB}) and (\ref{mmbulk}), we can finally calculate the correlation matrix:
\beq
{\cal G}={\cal M}{\cal B}{\cal A}^{-1}\,.\label{GG}
\eeq

Finally from the correlation matrix (\ref{GG}), it is straightforward to find the AC conductivity in superconducting phase,
using the Kubo formula:
\beq
\sigma(\omega)=\frac{{\cal G}_{11}}{i\omega}\,.
\eeq

In  figure \ref{ACplot} we plot the AC conductivity for the model (\ref{model2})
with the non-linear Lagrangian for the neutral scalars,
for $\Delta=2$, $q=4$, $\alpha\,u_h=\sqrt{2}$, $m^2L^2=0.025$. We consider various values
of temperature running from the normal phase to the superconducting phase. The AC conductivity for the normal phase of the model (\ref{model2})
has first appeared in \cite{Baggioli:2014roa}, where it has been shown
that (temperatures are measured in units of square root of charge density)\\

1. For $T>T_0$ ($T_0 \simeq 0.46$ (for the considered model) the system exhibits a metallic behavior, $d\sigma_{DC}/dT<0$.\\

2. For $T<T_0$ the system exhibits an insulating behavior, $d\sigma_{DC}/dT>0$.\\

3. For $T<T'<T_0$, where $T'\simeq 0.35$ (for the considered model) the non-trivial structure in
the AC conductivity appears. To be more precise a Mid-frequency \color{black} peak shows up signaling a weight transfer mechanism and an emerging collective degree of freedom.\\

These properties are illustrated in figure \ref{NormalFeat}. We checked that the sum rules for the optical conductivity are satisfied in both normal and broken phases.

After we couple this model of \cite{Baggioli:2014roa}, with the potential (\ref{model2})
for the neutral scalars, to the superconducting sector, more features appear. 
For the choice of parameters $\Delta=2$, $q=4$, $\alpha\,u_h=\sqrt{2}$, $m^2L^2=0.025$
we continue to enumerate what happens as we decrease the temperature:\\

4. At $T_c\simeq 0.16$ (for the considered model) the second order phase transition occurs.
The system lives in a superconducting phase, when $T<T_c$.\\

5. At $T=T''\simeq 0.1$ (for the considered model) the peak in the imaginary part of the
AC conductivity disappears. The peak in the real part of the AC conductivity in superconducting phase gets smaller
as the temperature is lowered and eventually disappears.\\

These properties can be seen in figure \ref{ACplot}.
We will comment more on these features in discussion section \ref{section7}.\\

It would be very interesting to find the QNM excitations of the system in both normal and broken phase to have complete control on its transport properties and its collective excitations. We leave this topic for future studies.
\color{black}

\section{Dome of superconductivity}\label{section6}

In this section we describe how to construct a superconducting
dome, by tuning the parameters of the model
(\ref{model2}) with the non-linear Lagrangian for the neutral scalars.
In nature, High-Tc superconductors exhibit a dome
of superconductivity (see figure~\ref{fig:DomeIntro}) between insulating and metallic normal phases as a function of dialing the doping of the sample. We will construct a qualitatively similar behaviour but increasing the disorder-strength of the system.\color{black}\\
Due to limitations of our system we cannot construct an actual insulator, however the non-linear model (\ref{model2})
still allows to distinguish between two qualitatively different states
of the normal phase, (\ref{metalder}) and (\ref{insulatorder}).

\begin{figure}
\centering
\includegraphics[width=.31\textwidth]{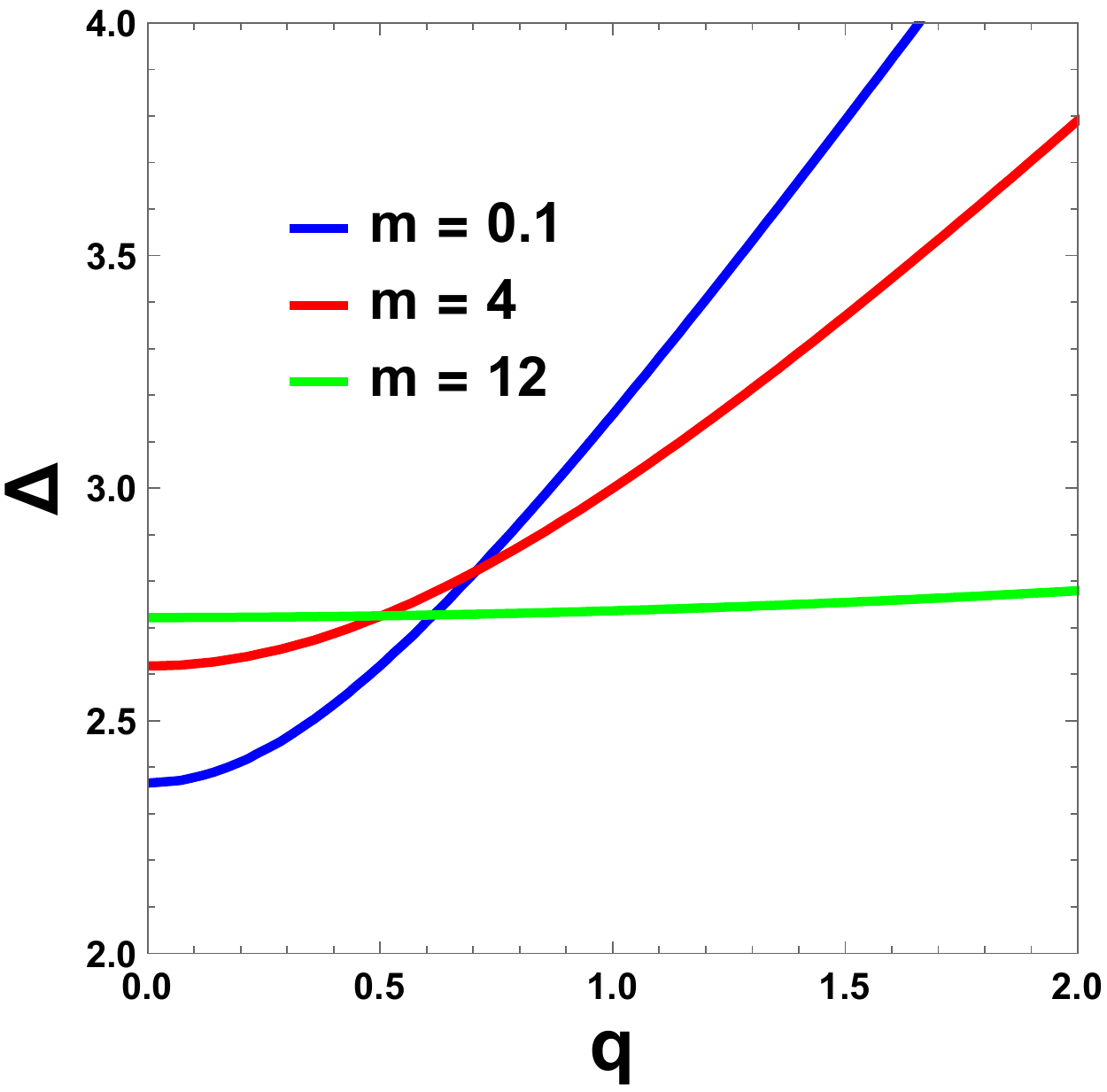}
\includegraphics[width=.31\textwidth]{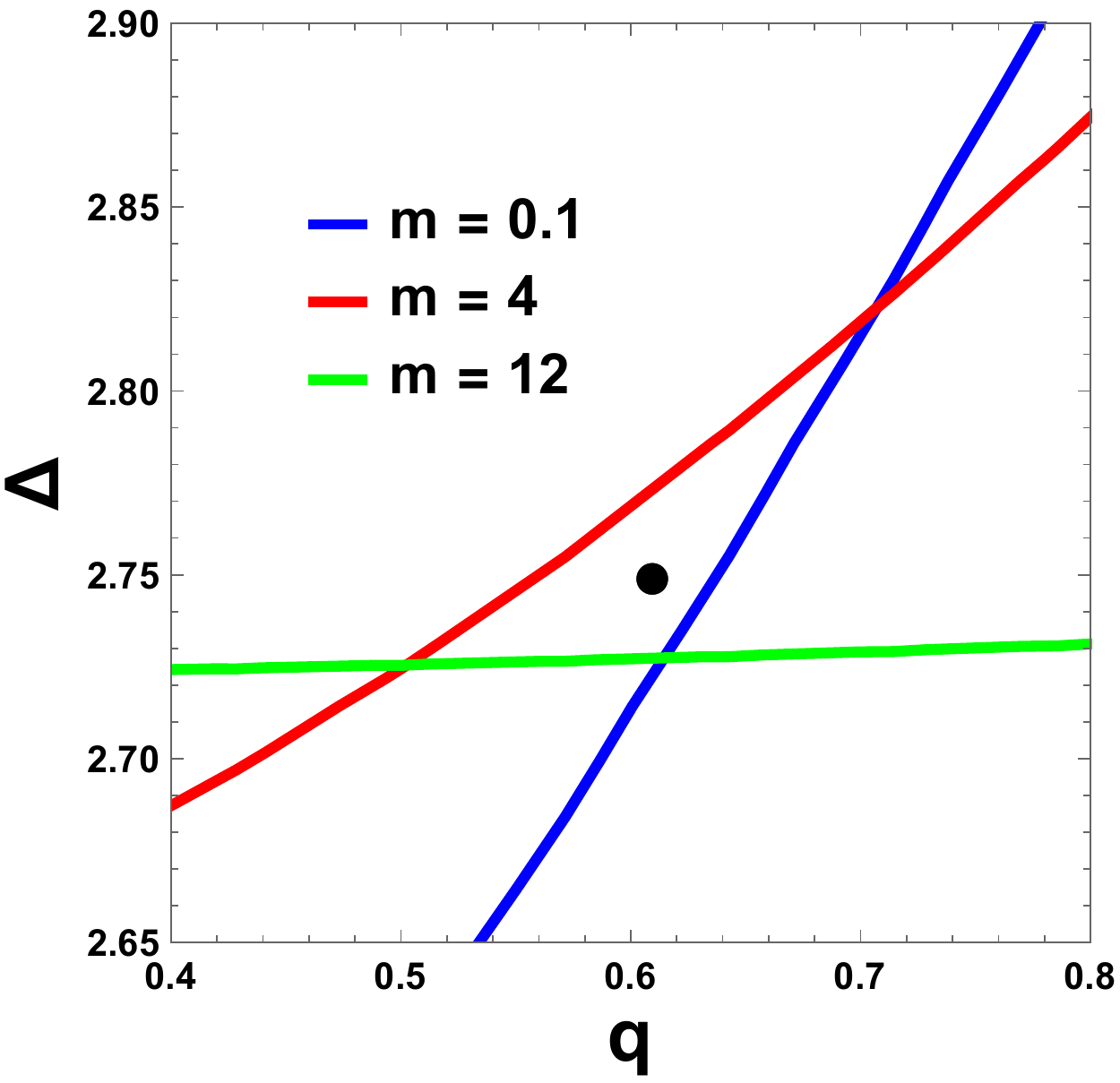}
\includegraphics[width=.33\textwidth]{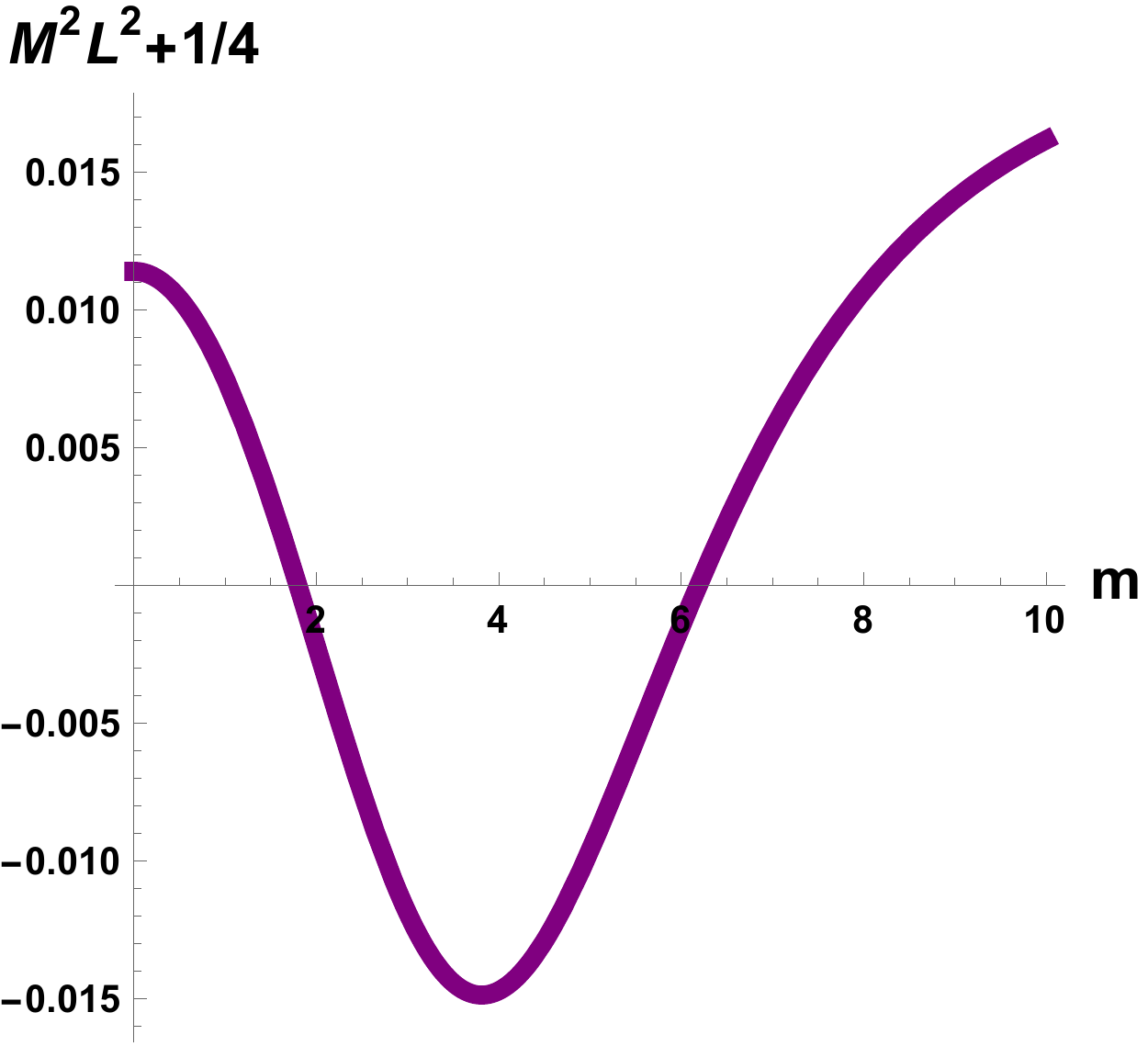}
\caption{Searching for the dome for the model (\ref{model2}) with $\alpha=0.25$
at vanishing temperature. From the left graph we conclude
that in order to have the superconducting dome we need to choose
$\Delta$ and $q$ from a small vicinity of the point (\ref{dmedeltaq}).
In the right graph we have verified explicitly the IR instability
of the model with $\Delta=2.75$, $q=0.61$,
between two finite values of $m$.}
\label{domeplot1}
\end{figure}

\begin{figure}
\centering
\includegraphics[width=.37\textwidth]{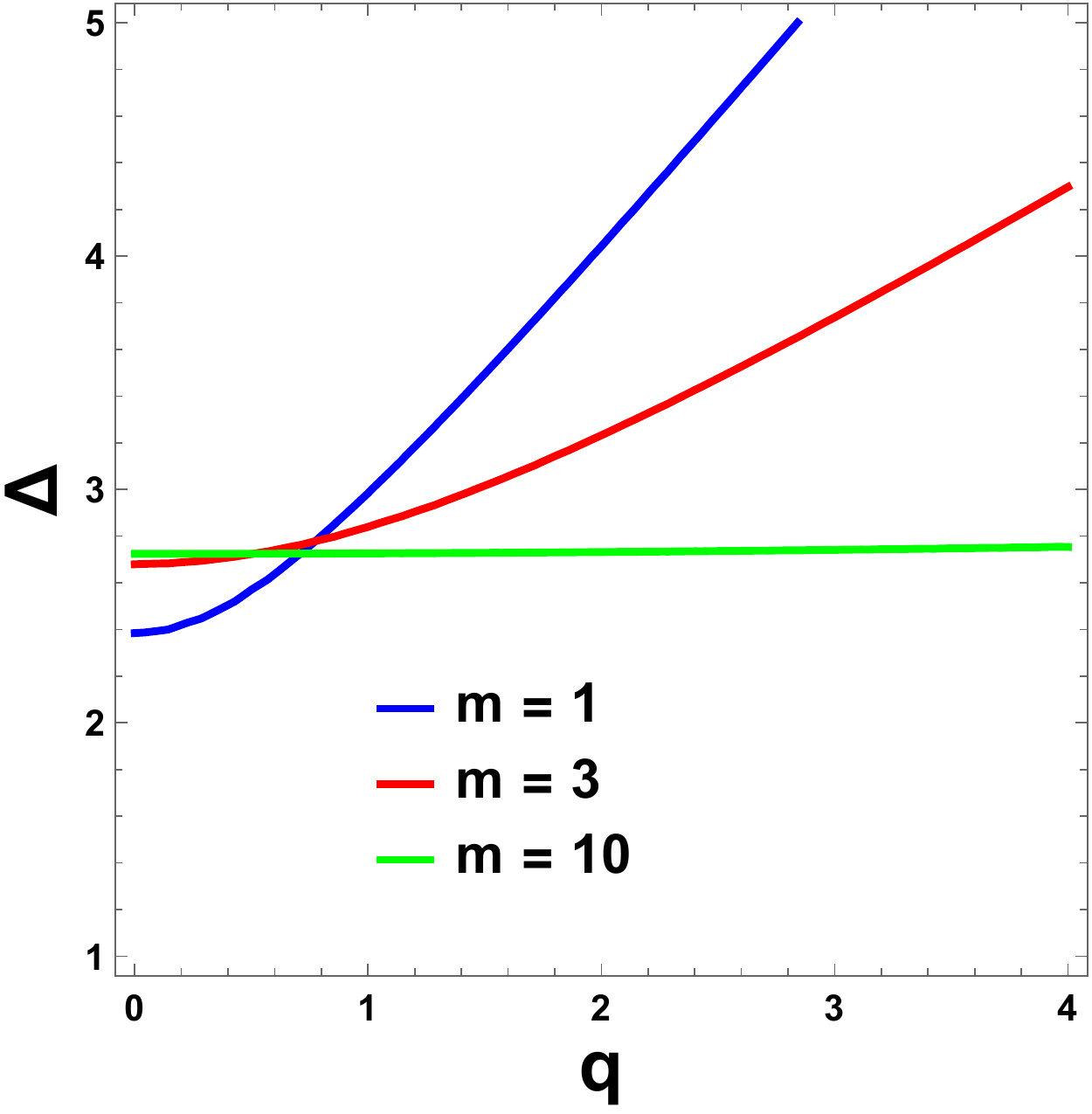}
\includegraphics[width=.4\textwidth]{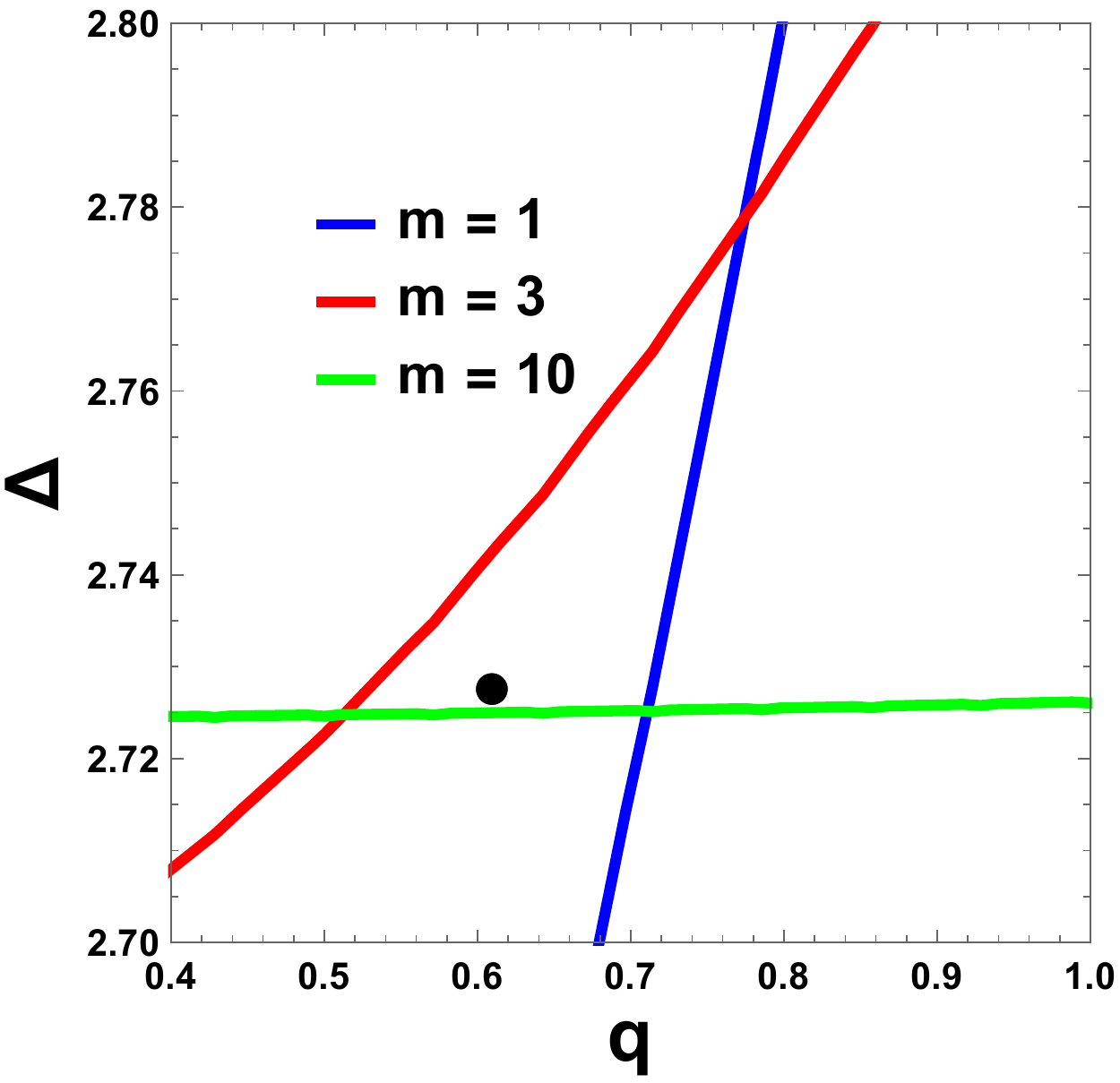}\\
\includegraphics[width=.4\textwidth]{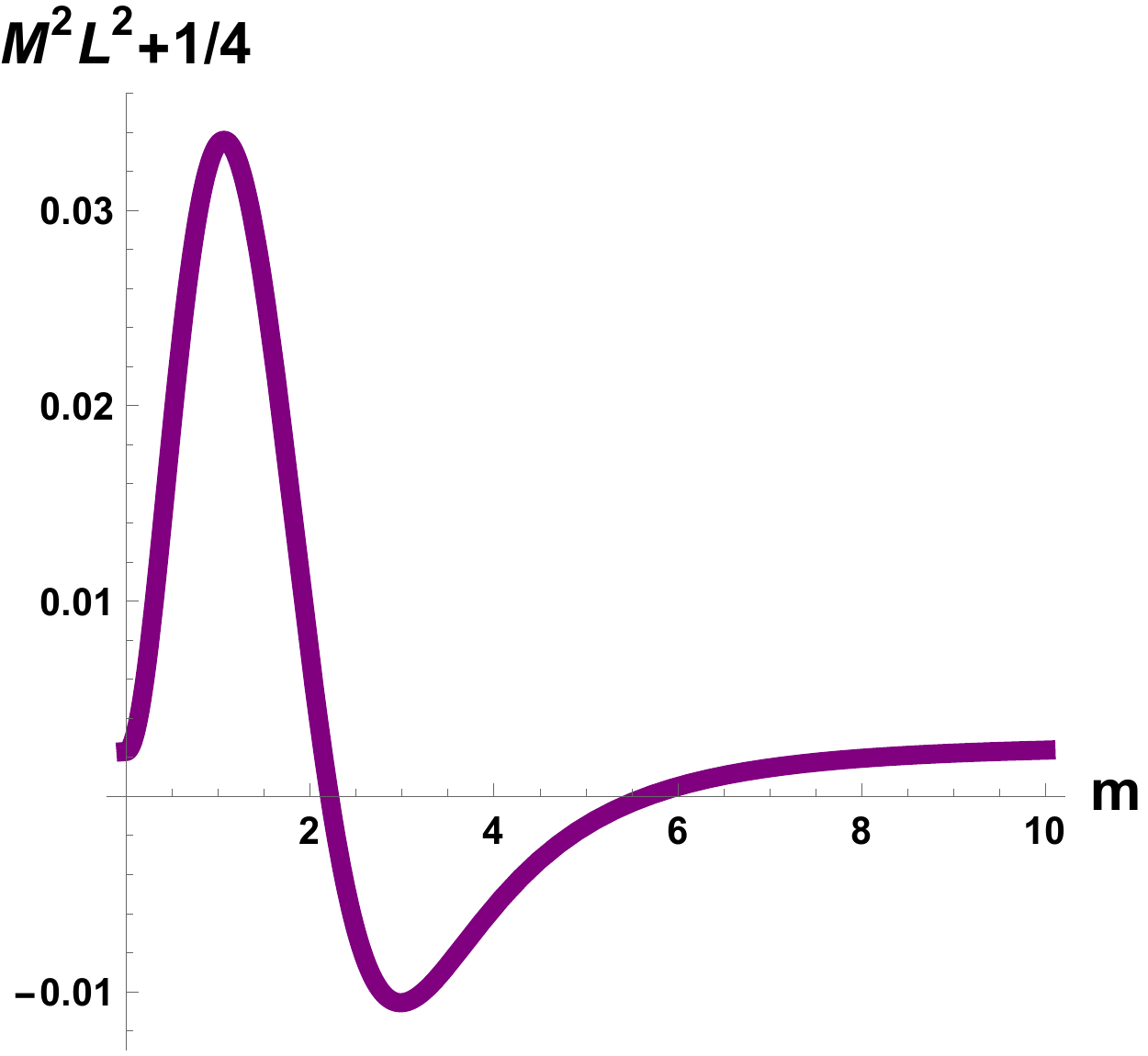}
\includegraphics[width=.4\textwidth]{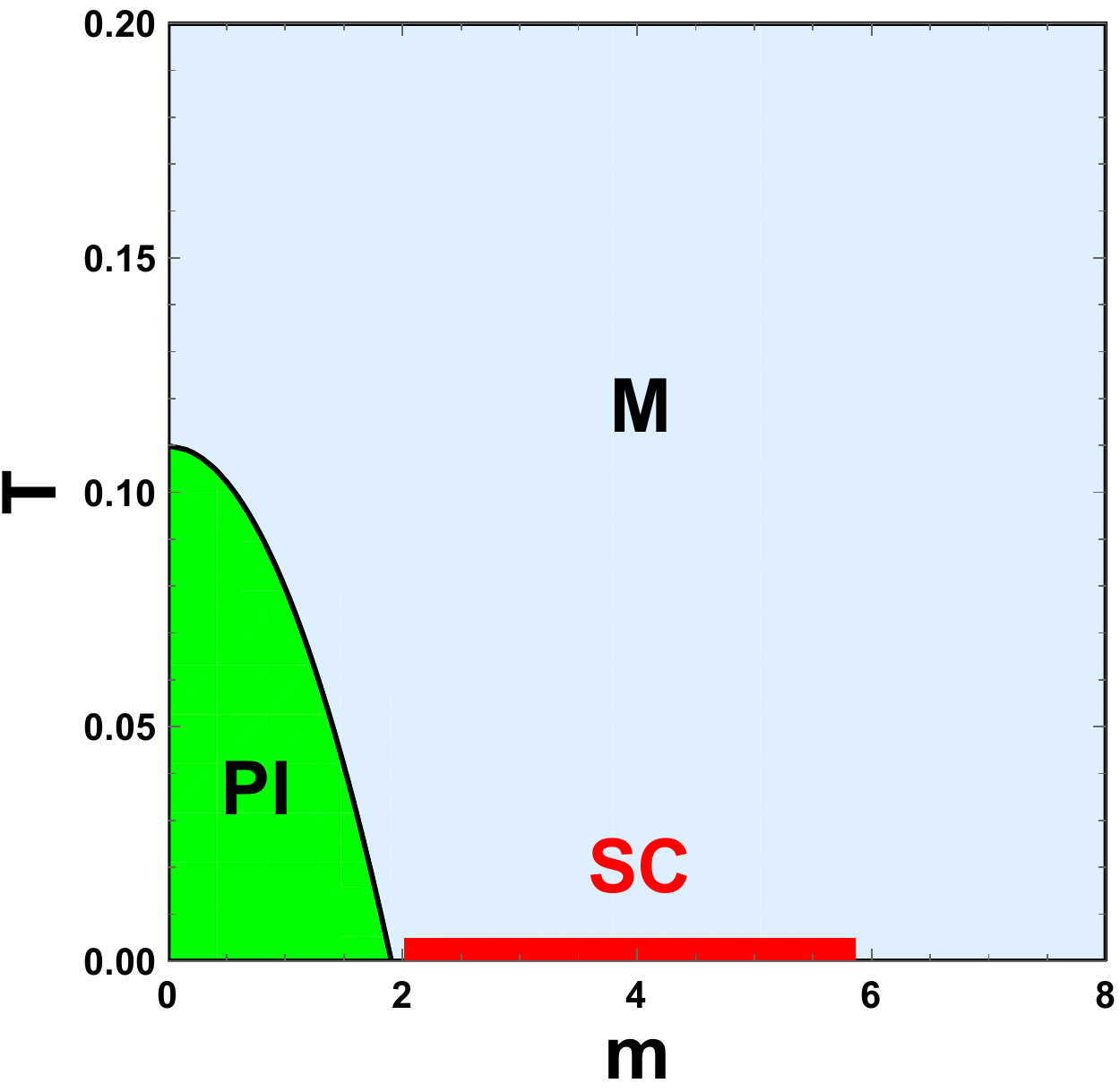}
\caption{\textbf{Top:} The $D=0$ contours for the model (\ref{model2}) with $\alpha=0.5$;
\textbf{Bottom Left:} BF bound violation for $\Delta=2.728$, $q=0.61$;
\textbf{Bottom Right:} Full Phase Diagram for the model. Green region
is a normal pseudo-insulating phase, grey region is a normal metallic phase,
red region is a superconducting phase.}
\label{domeplot2}
\end{figure}
The first observation is that when we decouple the translational-symmetry breaking sector
of the neutral scalar fields,
by setting $m=0$, we restore the framework of an ordinary holographic
superconductor. Therefore, in order to confine the superconducting
phase inside a dome, we need to make sure that
the ordinary holographic superconductor exists in the normal state at any temperature.
The way to achieve this is to make sure that the parameters
$\Delta$ and $q$ are such that the normal phase at $T=0$ is stable.
That is, we should have $D>0$, where $D$ is given by (\ref{Ddef}),
with $\kappa=0$ and $m=0$. The $T=0$ IR {\it stability} condition therefore reduces
to a well-known inequality, which reads:
\beq
3+2\Delta(\Delta-3)-4q^2>0\,.\label{IRstabregion}
\eeq

Suppose now we stay on top of the $T=0$ line on the $(\Delta,q)$
plane of ordinary holographic superconductor.
The next step in engineering a model 
exhibiting a superconducting dome, is to restore a superconductor at a finite value of $m=m_1$,
and then make sure that there is another value $m_2>m_1$, such that the system
at $m>m_2$ is again in a normal phase.
The procedure to search for the parameters which lead to
the superconducting dome is the following. For the chosen value of $\alpha$
we plot $D=0$ curves one the $(\Delta,q)$ plane, with the $D$ given by (\ref{Ddef}),
parametrized by various values of $m$.
We search for the points $(\Delta,q)$ of intersection of two curves, corresponding
to two different values of $m$. These values of $m$
can be the boundaries of the dome region at $T=0$.
We then verify this explicitly by plotting  the $D$ for given $\alpha$, $\Delta$, $q$.
In figure \ref{domeplot1} we plot the $D=0$ curves for $\alpha=0.25$, on the $(\Delta,q)$
plane, and demonstrate explicitly that the dome requirement
restricts us to consider a small sub-region on the $(\Delta,q)$ plane.
In figure \ref{domeplot2} we repeat this for $\alpha=0.5$, and also plot the corresponding phase diagram.
The superconducting phase is bounded from above by a small critical temperature,
and is represented on the graph by a red interval.

We have found that the requirement of having an interval of superconductivity
$[m_1,\,m_2]$ at $T=0$ is rather restrictive\footnote{
Being more specific, it seems that there exists a lower bound for $\Delta$ below which no SC dome can be built within this model. It would be nice to understand better this bound.}.
We have found that on order to achieve the `dome' at a vanishing
temperature we need to tune $\Delta$ and $q$ to a small subregion of the region
(\ref{IRstabregion}), centered around the point
\beq
(\Delta_d,\,q_d)\simeq (2.74,\,0.6)\,.\label{dmedeltaq}
\eeq
For such $\Delta$ and $q$ we can engineer a model which,
at $T=0$, exists in a normal pseudo-insulating phase for $m\in [0,\, m_1]$, in a superconducting phase
for $m\in [m_1,\,m_2]$, and a normal metallic phase for $m>m_2$.

The next step to construct the superconducting dome
is to study the phase structure of the system at finite temperature.
To determine the boundary of the superconducting region, that is the line
of the second oder superconducting phase transition,
we can start in the normal phase, at larger values of temperature,
and determine when it becomes unstable towards formation
of the scalar hair. This procedure has been reviewed in Subsection \ref{subfininst}.

However, the point (\ref{dmedeltaq}) is very close to the 
boundary of the $T=0$ infrared instability region of the model (\ref{model2}). This behavior is rather generic and leads to the conclusion that the height size of the dome is very limited, the $T_c$ is very small and not accessible through stable numerical analysis. Another way of realizing this issue relies on noticing that the BF bound is very mildly violated in the dome region such that the instability is very soft.

\begin{figure}
\centering
\includegraphics[width=.37\textwidth]{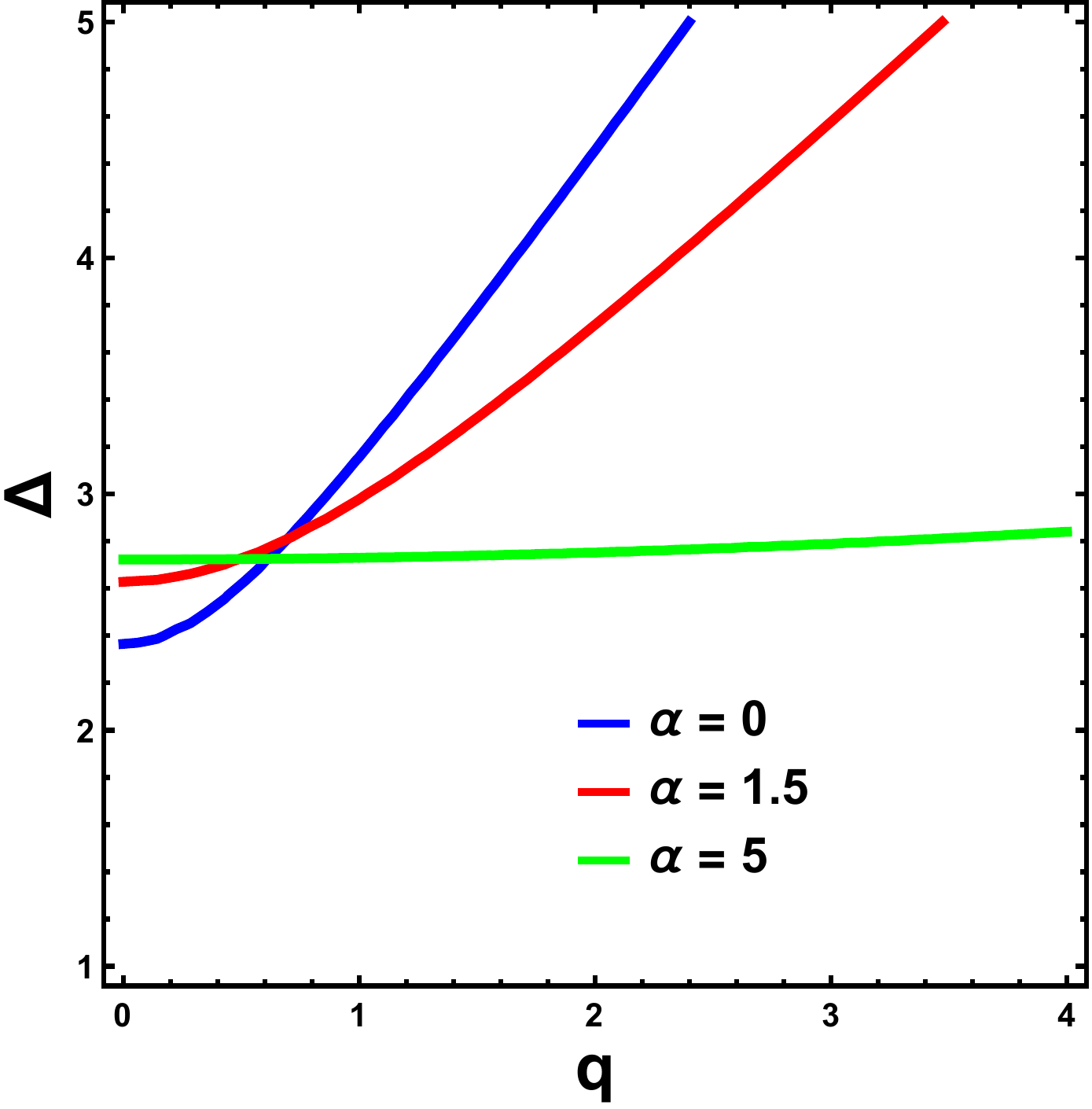}
\includegraphics[width=.4\textwidth]{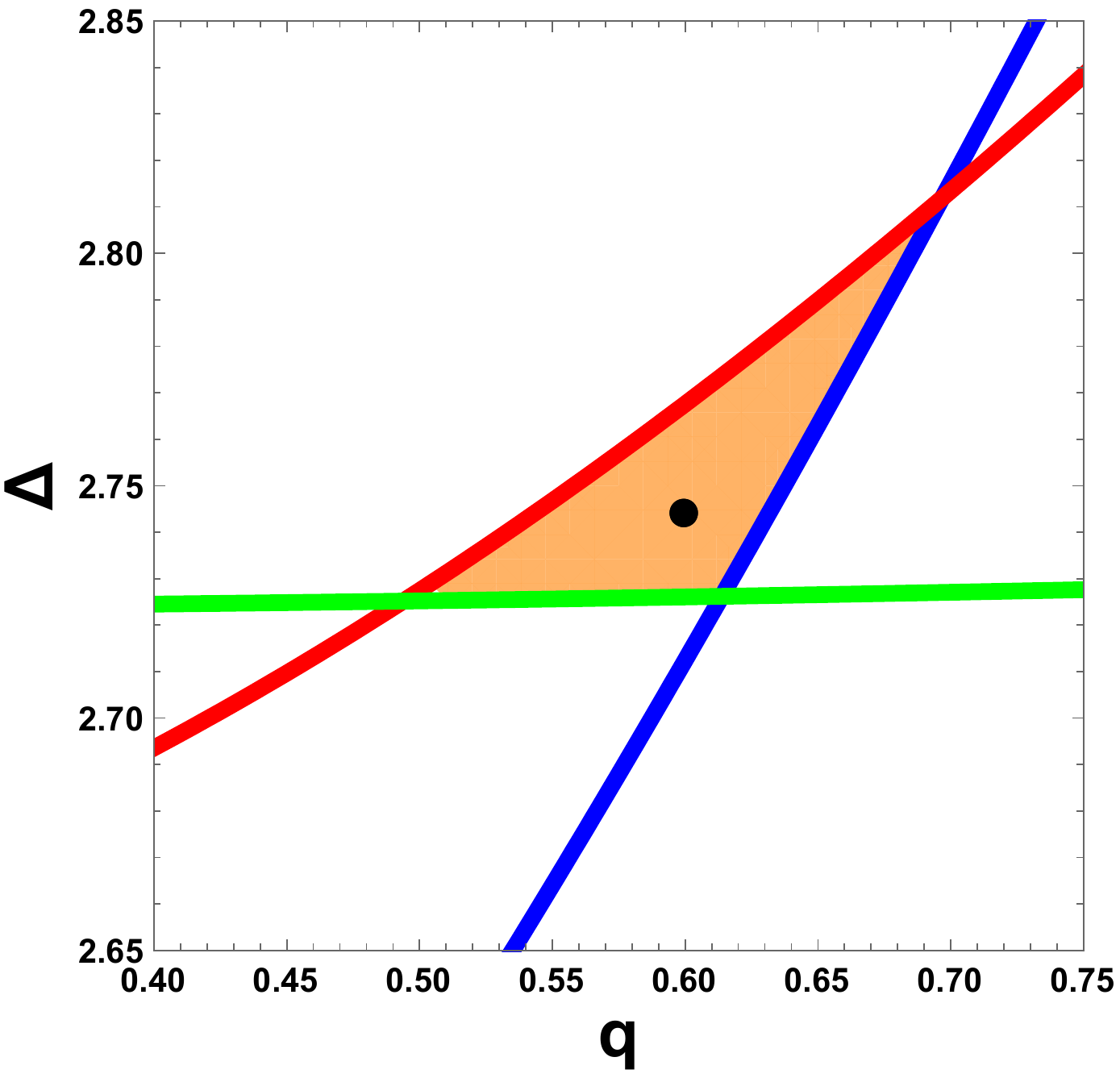}\\
\includegraphics[width=.37\textwidth]{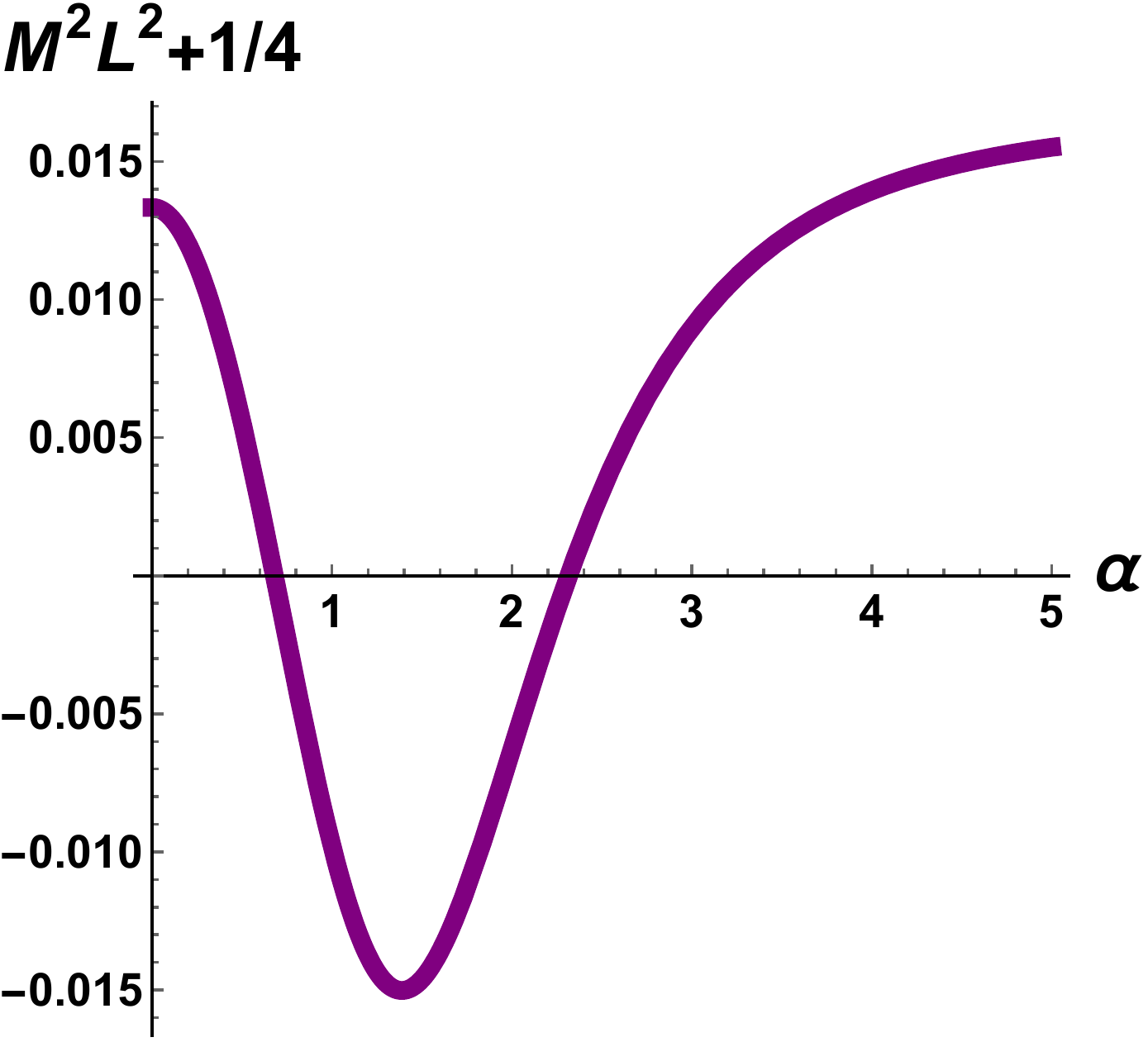}
\includegraphics[width=.4\textwidth]{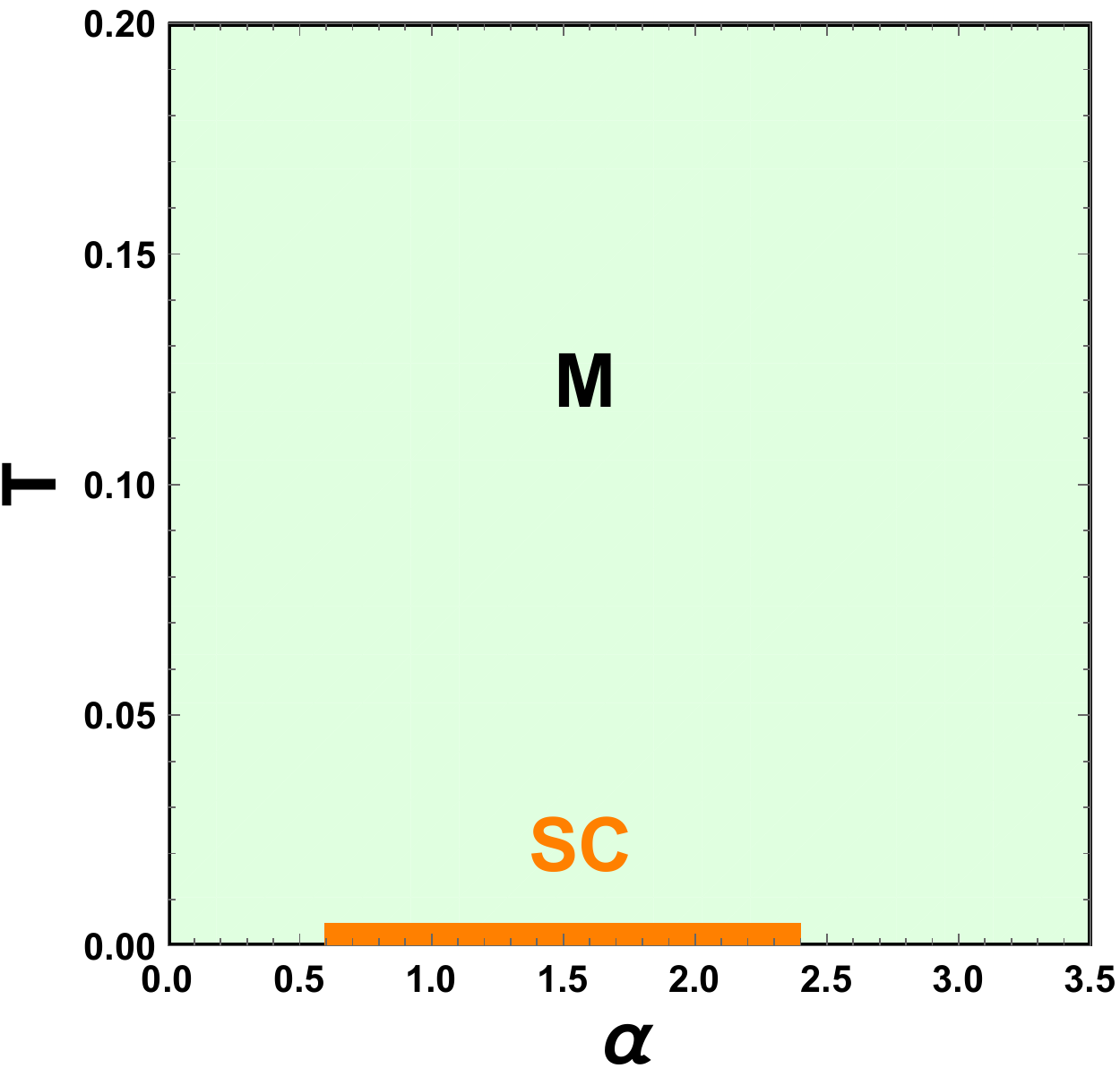}
\caption{
\textbf{Top:} The $D=0$ contours for the model (\ref{model1});
\textbf{Bottom Left:} BF bound violation for $\Delta=2.745$, $q=0.6$;
\textbf{Bottom Right:} Full Phase Diagram for the model. Orange region is a superconductor, lettuce region is a metal.}
\label{domeplotlin}
\end{figure}

We have repeated a similar dome analyses for the model (\ref{model1}),
with the parameter $A=\alpha \,m$ playing the role of disorder-strength in the boundary theory\color{black}.
Interestingly enough, we have found that again for $\Delta$ and $q$
tuned to a small vicinity of the point (\ref{dmedeltaq}) we obtain
a superconducting dome. This time, however, the normal
phase can only be metallic. The superconducting dome at $T=0$
is an interval $[A_1,\,A_2]$, existing between two regions
of normal metallic state, at $A\in [0,\,A_1]$ and $A>A_2$.
The critical temperature is bounded from above by a small number, 
and we did not access the finite-temperature superconducting state.
We plot our results for the dome in model (\ref{model1}) in figure \ref{domeplotlin}.

This analysis shows that the existence of a superconducting dome region is a rather generic feature of these models, independent of the choice of the potential.
In the next section we discuss the possible ways to alleviate
the problem of the flatness of the dome. This seems to require
introduction of an extra elements to our holographic system.

\section{Discussion}\label{section7}

In this paper we considered a holographic
superconductor with broken translational symmetry, continuing the
research, initiated in \cite{Andrade:2014xca,Kim:2015dna}.
To break the translational symmetry we used the known
technique \cite{Andrade:2013gsa}, coupling our system to the sector of massless neutral
scalar fields, depending linearly on the spatial coordinates.
We studied the standard Lagrangian for these neutral scalars,
as well as its non-linear generalization, proposed in \cite{Baggioli:2014roa}.\\
We have constructed models, exhibiting the following non-trivial new
features:
\begin{enumerate}
\item
The Holographic superconductor in the non-linear Lagrangian model has a rich phase diagram on the temperature-disorder strength plane.
In particular the superconducting phase is separated from the normal
pseudo-insulating phase and the normal metallic phase by the line
of second order phase transition as shown in figure \ref{PhaseDPlot}.
\item
In the same model the optical conductivity exhibits a non-trivial emerging structure,
signaling a collective excitation of the charge carriers localized in the mid-frequency range\color{black}.
This has been observed in \cite{Baggioli:2014roa} in the normal
phase of the same model, for temperatures, lower than a certain critical value.
In this paper we have demonstrated that this structure persists in the superconducting phase.
Eventually it gets destroyed by the charge condensate. This suggests a possible competition between the superconducting mechanism and the momentum dissipating one. In particular it seems clear that a large superfluid density completely screens this collective excitation which in a sense gets eaten by the large condensate. We are not aware of real superconducting system supporting a collective localized excitation like the one we see. In \cite{Baggioli:2014roa} this excitation was compared to a polaron excitation; it would be definitely interesting to make a comparison between the behaviour of this collective excitation in the holographic model and what really happens to a polaron when superconductivity onsets. Unfortunately we are not aware of such a mechanism in real condensed matter systems.
 \color{black} It would be also nice to see if other holographic models, providing translational symmetry breaking, support the same property. In this direction it would be very interesting to study the QNM structure of the system as initiated in \cite{Davison:2014lua}.\color{black}
\item We performed a complete analysis of the behavior of
the critical (superconducting) temperature as a function of the various parameters of our model. In particular we  studied the curious non-monotonic behavour of $T_c$ as a function of the graviton mass $m$, which was already observed in \cite{Kim:2015dna,Andrade:2014xca}. Our results suggest that this feature persists for generic Lagrangian for neutral scalars. We do not have any clear explanation of the big mass regime where $T_c$ actually increases with the strength of translational symmetry breaking. It is even tempting to doubt the  model in that regime, reminiscing the following known issues: for large momentum dissipation it seems that the energy density of the dual field theory at zero charge density gets negative \cite{Davison:2014lua}; the diffusion bounds for the model are unrestricted from below and the diffusion constants go to zero in that limit \cite{Amoretti:2014ola}.

A very similar behavior has been observed in holographic SC with helical lattices \cite{Erdmenger:2015qqa} and  with disorder \cite{Arean:2014oaa}. It would be interesting to further analyze the universality and the meaning of this feature.\color{black}
\item
By tuning the values of the scaling dimension $\Delta$ and the charge $q$
of the scalar field, which is a bulk dual to the charge condensate of the boundary superconductor,
one can obtain a system, which exists in a superconducting phase, enclosed in a dome region. The dome region occurs upon increasing the disorder-strength parameter of the model, behaviour which is definitely different from the actual High-Tc SC phase diagram where this happens because of the doping of the material\color{black}.
The critical temperature of the dome is very small, and in fact appears
to be too hard to calculate numerically. The superconducting dome exists
for both linear and non-linear models. In the case of the model
with the linear Lagrangian for the neutral scalars, the superconducting dome
exists in the middle of the normal metallic phase.
In its non-linear extension instead, the dome exists between
a pseudo-insulating phase for smaller values of disorder strength, and a metallic
phase for larger values. There are no experimental evidences of dome regions occurring because of disorder. We hope our work could in a way motivate some experimental effort in that direction.\color{black}

We are aware of two only holographic examples which show a superconducting dome region in way different setups \cite{Ganguli:2013oya,Gauntlett:2009bh}. \color{black}
\end{enumerate}
It would be interesting to improve the model, so that pseudo-insulating
phase is replaced by an actual insulating phase. This will make the phase diagram more resembling
such of an actual high-Tc superconductor. This could be easily achieved introducing a dilaton
field into the model.

One direct expectation of our Holographic model is that the dome of superconductivity we find seems to only exist in a very fine tuned region of parameter space which is always very close to the zero temperature instability. As a conseguence two immediate questions arise:
\begin{itemize}
\item Is it possible to enlarge significantly the region of the parameter space where th dome appears?
\item Is it possible to get a dome with a reasonable $T_c$ which can be numerically be resolved?
\end{itemize}
Solving the second issue would be indeed very important to rule out possible non-IR instabilities that would remove the dome region.\\
One of the ways to accomplish this might be realized by the inclusion of a non-trivial coupling $\kappa$
between the charged scalar condensate and the neutral scalars as already shown in the action \ref{Iact}. Generically it seems that without the introduction of additional elements this can not be obtained. We leave this question for future work.\color{black}

A further interesting question is to look at universal properties of these large class of effective toy models such as the accomplishment of Homes' Law following \cite{Erdmenger:2012ik}.
We leave these interesting questions for future investigation.\color{black}

\section*{Acknowledgements}

We would like to thank Richard~Davison, Daniel~Arean, Siavash~Golkar, Gary~Horowitz, Keun-Young~Kim, Rene~Meyer, Eun-Gook~Moon, Nick~Poovuttikul
and Matthew~Roberts for valuable discussions and comments. We would like to thank
Oriol~Pujol\'as for initial collaboration on the project and for insights about the superconducting dome. We would also like to thank the anonymous referees for valuable comments and suggestions.\color{black}
\hspace{0.07cm}MB acknowledges support from MINECO under grant FPA2011-25948, DURSI under grant 2014SGR1450 and Centro de Excelencia Severo Ochoa program, grant SEV-2012-
0234. The work of MG was supported by Oehme Fellowship.
\appendix

\section{Condensate and grand potential}\label{appendix2}
The aim of this appendix is to provide more details about the computations and the numerical procedures we did in Section \ref{section4}.\color{black}
\subsection{Condensate}

In this subsection we will outline the routine to obtain the numerical solution of the equations of motion
(\ref{sEinst})-(\ref{scfeq}) for the whole superconducting background.
First of all, evaluating the equations (\ref{sEinst})-(\ref{scfeq}) at $u=u_h$,
we can express $\psi'(u_h)$, $f'(u_h)$, $\chi'(u_h)$, $A_t''(u_h)$
in terms of $\psi(u_h)$, $\chi(u_h)$, $A_t'(u_h)$.
Therefore we impose the initial conditions at $u_h-\epsilon$ in the following way:
\begin{align}
\psi(u_h-\epsilon)&=\psi(u_h)-\epsilon \psi'(u_h)\,,\quad \psi'(u_h-\epsilon)=\psi'(u_h)\,,\notag\\
f(u_h-\epsilon)&=-\epsilon f'(u_h)\,,\quad \chi (u_h-\epsilon)=\chi(u_h)-\epsilon \chi'(u_h)\,,\label{nhbcsc}\\
A_t(u_h-\epsilon)&=-\epsilon A_t'(u_h)-\frac{\epsilon^2}{2}A_t''(u_h)\,,
\quad A_t'(u_h-\epsilon) =A_t'(u_h)-\epsilon A_t''(u_h)\,.\notag
\end{align}
where $\epsilon$ is a small IR cutoff. One can solve equations of motion near the horizon
to arbitrary order in $\epsilon$. We have found that imposing (\ref{nhbcsc}) is sufficient.
We have checked explicitly that the results are stable towards changing $\epsilon$.
We have the freedom of choice of the initial conditions $\psi(u_h)$, $A_t'(u_h)$,
and $\chi(u_h)$. The freedom of choice of $\chi(u_h)$
is spurious, due to the time scaling symmetry, as we discuss below.

\begin{figure}
\begin{center}
\includegraphics[width=.55\textwidth]{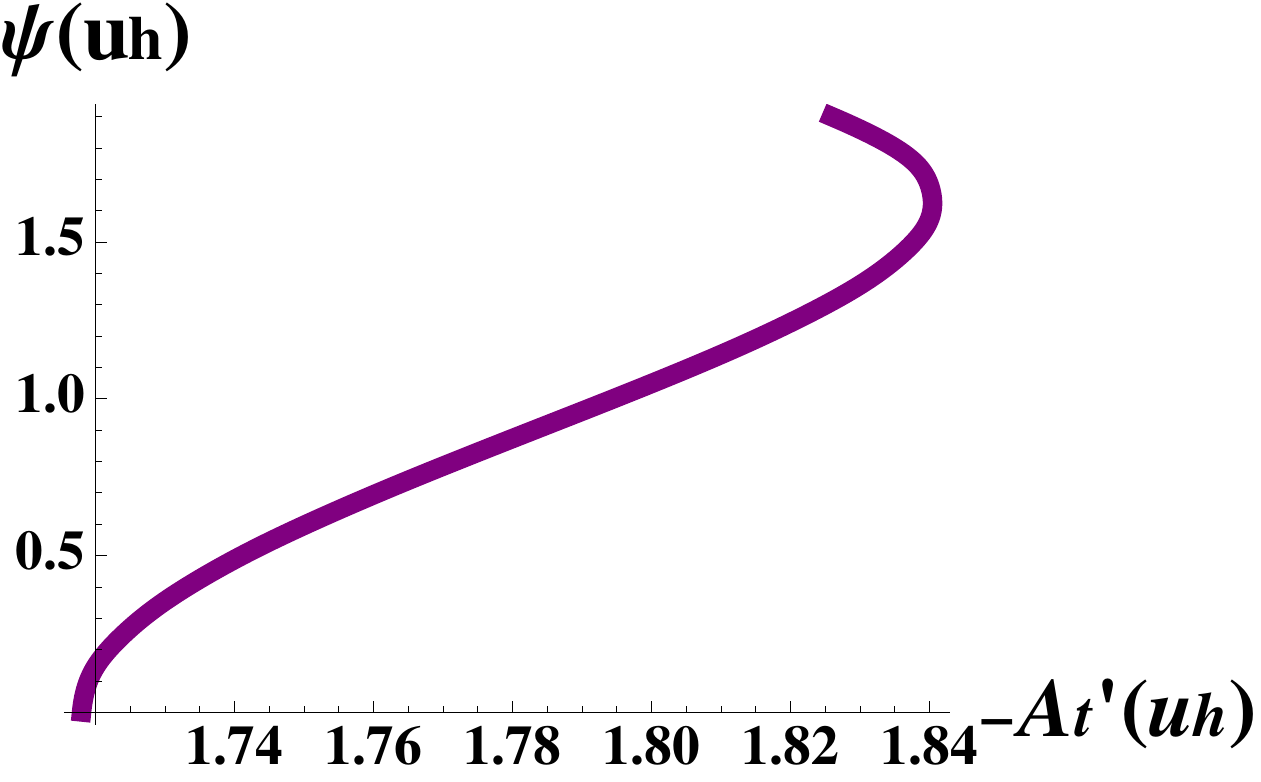}
\end{center}
\caption{Condensate for $\Delta=2$, $q=1$, $\alpha\,u_h=0.5$, $m\,L=1$ model
with $V=z+z^5$, and the corresponding imposed near-horizon data (for $\chi(u_h-\epsilon)=1$).}
\label{horcond}
\end{figure}

The values of $\psi(u_h)$ and $A_t'(u_h)$ are fixed by the requirement of having
a fixed temperature $T/\rho^{1/2}$ and zero source $\psi_1=0$, see (\ref{nbexp}).
Both the charge density $\rho$, in units of which me measure the temperature, and the source $\psi_1$
are determined by the near-boundary behavior of the numerical solution, with the gauge field behaving as:
\beq
A_t(u)=\mu-\rho \,u+{\cal O}(u^2)\,.
\eeq

In practical calculation we do the following.
Suppose the temperature is sufficiently small, so that the system is in a superconducting phase.
We know that increasing the temperature will decrease the condensate, $\psi_2/\rho^{\Delta/2}$,
until finally at the critical temperature $T_c$ the condensate is zero.
At that point $\psi(u_h)=0$, that is, we do not have the solution with vanishing source and non-trivial
profile of $\psi(u)$ in the bulk.
Therefore we can start at $\psi(u_h)=0$, and take gradually incrementing values of $\psi(u_h)$.
For each value of $\psi(u_h)$ we search for $A_t'(u_h)$, such that $\psi_1=0$. For an example of this kind of result see figure \ref{horcond}.
Finally for the given pair $(\psi(u_h),A_t'(u_h))$ we calculate numerically $\left(T/\rho^{1/2},
\psi_2/\rho^{\Delta/2}\right)$ as shown for example in figure \ref{fig:ConductivityNeutralk=0}.\\

{\it Scaling symmetry}\\

The equations of motion (\ref{sEinst})-(\ref{scfeq}) are invariant under the scaling symmetry:
\beq
u={\tilde u}/a\,,\quad (t,x,y)=({\tilde t},{\tilde x},{\tilde y})/a\,,\quad A_t={\tilde A_t}\,a\,,\quad \alpha={\tilde \alpha}\,a\,,\label{scsym}
\eeq
where $a$ is a parameter of the symmetry transformation. The temperature, chemical
potential, and the charge density therefore transform as:
\beq
T=a\,\tilde T\,,\quad \mu=a\,\tilde \mu\,,\quad \rho=a^2\,\tilde \rho\,.
\eeq
The scaling symmetry (\ref{scsym}) allows one to fix $u_h=1$.
If $u_h$ is not fixed to one, then we should substitute $u_h^{-2}\tilde A_t'(u_h)$
as the initial condition for the flux at the horizon. We have checked explicitly
that the results are invariant under change of $u_h$.\\

{\it Time scaling symmetry}\\

The equations of motion (\ref{sEinst})-(\ref{scfeq}) are invariant under the
time scaling symmetry:
\beq
e^\chi=b^2\,e^{\tilde \chi}\,,\quad t=b\,\tilde t\,,\quad A_t=\tilde A_t/b\,,\label{tscsym}
\eeq
where $b$ is a parameter of the symmetry transformation.

We can use the time scaling symmetry (\ref{tscsym})
to fix $\chi(0)=0$ at the boundary. This is necessary, so that the
speed of light in the boundary field theory is equal to one. To achieve this, we impose
the initial conditions on $\chi$ to be $\chi(u_h)+2\log\,b$,
and on the flux to be $\tilde A_t'(u_h)/b$. We fix $\chi(u_h)$
once and for all. We have demonstrated explicitly that the result
is independent of the choice of $\chi(u_h)$.

After fixing $\chi(u_h)$,
for the given $\tilde A_t'(u_h)$, we integrate numerically the equations
of motion, with $b=1$. We then impose $b=e^{-\chi(0)/2}$,
where $\chi(0)$ is determined numerically. For this $b$
we impose the initial conditions $\chi(u_h)+2\log\,b$, $\tilde A_t'(u_h)/b$ and 
integrate the equations of motion again. This time, due to the time
scaling symmetry (\ref{tscsym}), we have $\chi(0)=0$.
We have verified this explicitly.

Running the described numerical procedure we were able to construct the condensate
$\psi_2/\rho^{\Delta/2}$ as a function of temperature $T/\rho^{1/2}$.
In figure \ref{fig:ConductivityNeutralk=0} we provide the plot of the condensate, for the model (\ref{model2}) with $\Delta=2$,
$q=1$, $\alpha\,u_h=0.5$ ($\alpha$ in units of entropy density), $m\,L=1$.
For the same parameters we also plot
the initial conditions $(-A_t'(u_h),\psi(u_h))$, which we imposed, to enable the
vanishing source $\psi_1=0$ in figure \ref{horcond}.

\subsection{Grand potential}

Here we provide intermediate steps for calculation of the grand potential.

The holographic prescription for the calculation of the grand potential is:
\begin{equation}
\Omega\,=\,-T\,\log Z\,=\,T \mathcal{S}_{E}\,,
\end{equation}
where $\mathcal{S}_{E}$ is a Euclidean on-shell action of the bulk theory.
This should be supplemented with the boundary Gibbons-Hawking term,
and the counter-terms\footnote{Equivalently, one can calculate difference
of the grand potentials of two phases, in order to avoid adding the counter-terms.}.
The resulting action reads:
\beq
\mathcal{S}_{E}=I+I_{GH}+I_{c.t.}\,,
\eeq
where the boundary Gibbons-Hawking term is given by:
\beq
I_{GH}=-2\int d^3x\,\sqrt{-h}\,K\,\Bigg|_{u=\epsilon}\,,\label{GHdef}
\eeq
where $\epsilon$ is a UV cutoff, $h_{ab}$ the pullback metric on the boundary and $K_{ab}$ the extrinsic curvature\footnote{It is defined by \beq
\label{extrcurv}
K=\nabla_\mu n^\mu\,,\qquad n^\mu=\left(0\,,\;0\,,\;0\,,\;u\,f(u)^{1/2}/L\right)\,.
\eeq
where $n^\mu$ is the unit vector normal to the boundary.}.
The counter-term action $I_{c.t.}$ is a sum of gravitational, scalar and axion fields counter-terms \cite{Balasubramanian:1999re,Andrade:2014xca}:
\beq
I_{c.t.}=-\int d^3x\,\sqrt{-h}\,\left[\frac{4}{L}+\frac{1}{L}\psi^2-2m^2\,L\,V\right]_{u=\epsilon}\,.\label{ctact}
\eeq
It is convenient to evaluate the following Lagrangian on shell:
\beq
\label{GHder}
\tilde L=L+2\,\left(\sqrt{-h}\,K\right)'\,.
\eeq
to get:
\beq
I+I_{GH}=\int d^4x\,\tilde L-2\int d^3x\,\sqrt{-h}\,K\Bigg|_{u=u_H}\,.\label{ItGHspl}
\eeq
After a straightforward calculation we obtain:
\beq
\label{Ltotder}
\tilde L=B'\,,\qquad B=L^2\left(A_tA_t'e^{\chi/2}-4u^{-3}f\,e^{-\chi/2}\right)\,.
\eeq
Notice that which $B(u_H)=0$.
Therefore the full on-shell action is given by:
\beq
I+I_{GH}=\int d^3x\,\left(\frac{4\pi L^2 T}{u_H^2}-B(\epsilon)\right)\,.\label{BrH}
\eeq
To proceed with the calculation, we need to be able to evaluate the counter-term action 
(\ref{ctact}) and the $B(\epsilon)$ term of (\ref{BrH}).
We need to know the
near-boundary behavior of the fields. That is given by
\footnote{Note that this is true only if the potential reads $V(X)=X+X^{n_1}+X^{n_2}+...$ where the smallest power is always equal to one.} :
\begin{align}
\label{Uvas}
A_t&=\mu-\rho u+\OO(u^2)\\
\psi&=\frac{\psi_1}{L^{3-\Delta}}\,u^{3-\Delta}+\frac{\psi_2}{L^\Delta}u^\Delta+\OO(u^{\Delta+1})\\
f&=1+\gamma_1 u^2+\gamma_2 u^3+\OO(u^4)\label{nbbas}\\
\chi&=\zeta_1 u^2+\zeta_2 u^3+\OO(u^4)\\
V&=V_1u^2+\OO(u^3)\,.
\end{align}
We are interested in the systems with vanishing source of the charged scalar, $\psi_1=0$.
By solving equations of motion near the boundary, we obtain:
\beq
\gamma_1=-m^2\,L^2\,V_1\,.
\eeq
Combining all the results together, we arrive at the final expression for the on-shell action:
\beq
\mathcal{S}_{E}=\int d^3x\,\left(16\pi\,S\,T+2L^2\gamma_2+L^2\mu\rho\right)\,.
\label{onsht}
\eeq

\section{On-shell action for fluctuations }\label{appendix1}

The calculation of the on-shell action for fluctuations is similar to the one for 
the grand potential performed in appendix \ref{appendix2}. 
The total action is a sum of the total bulk action (\ref{Itotb}), the Gibbon-Hawking (GH)
term on the boundary (\ref{GHdef}), and the counter-term action (\ref{ctact}):
\beq
I^{f}_{tot}=I^{f}_b+I^f_{GH}+I^f_{c.t.}\label{flact}
\eeq
We evaluate the action (\ref{flact}) on the ansatz:
\begin{align}
&ds^2=\frac{L^2}{u^2}\left(-f(u)dt^2+2\epsilon h_{tx}(u,t)dtdx+dx^2+dy^2+\frac{1}{f(u)}du^2\right)\,,\notag\\
&A=A_tdt+\epsilon\, a_x(u,t)dx\,,\notag\\
&\phi^x=\alpha\, x+\epsilon\, \xi (u,t)\,,\label{flans}\\
&\phi^y=\alpha \,y\,,\notag\\
&\psi=\psi(u)\,,\notag
\end{align}
and collect $\OO(\epsilon^2)$ terms, which describe dynamics of the fluctuations
$h_{tx}$, $a_x$, $\zeta$. The $\OO(\epsilon)$ terms vanish due to equations of motion,
satisfied by the background fields $f$, $A_t$, $\phi^{x,y}$, $\psi$,
and the $\OO(\epsilon^0)$ terms are contributions to the grand potential for the background.

The GH term vanishes at the horizon. Therefore:
\beq
\tilde I^f=I^f_{b}+I^f_{GH}=I^f_b-\left(-2\sqrt{-h}\,K\right)'\,.
\eeq
We obtain:
\begin{align}
\tilde I^f&=\frac{L^2\,e^{-\frac{\chi}{2}}}{4u^4\,f^2}\left(
-2 f e^{\chi } \left(2 u h_{tx}(u,t) \left(u^3 f A_t'\p_u a_x(u,t)+2 q^2 u A_t \psi ^2 a_x(u,t)\right.\right.\right.\notag\\
&+\left.\left.\left.4 f\p_u h_{tx}(u,t)+2 \alpha  L^2 m^2 u
\p_t\xi (u,t)\dot V\right)+u^2 \left(-\left(u^2 (\p_t a_x(u,t))^2\right.\right.\right.\right.\notag\\
&+\left.\left.\left.\left.f(\p_t h_{tx}(u,t))^2+2 L^2 m^2(\p_t \xi (u,t))
^2\dot V\right)\right)+h_{tx}(u,t)^2 \left(-2 \left(u f'+3\right)\right.\right.\right.\notag\\
&+\left.\left.\left. f \left(u^2 \psi ^{\prime 2}+2 u \chi '-6\right)+L^2 \left(2 m^2
\left(V-\alpha ^2 u^2 V'\right)+M^2 \psi ^2\right)\right)\right)\right.\notag\\
&-\left. u^2 e^{2 \chi} h_{tx}(u,t)^2 \left(u^2  A_t^{\prime 2}
+2 q^2 A_t^2 \psi^2\right)-2 u^4 f^3 (\p_u a_x(u,t))^2\right.\notag\\
&-\left. 4 q^2 u^2 f^2 \psi ^2 a_x(u,t)^2-4 L^2 m^2 u^2 f^3
(\p_u \xi (u,t))^2 \dot V\right)
\end{align}
To proceed, we integrate the $a_x^{\prime 2}$, $h_{tx}^{\prime 2}$, $\xi^{\prime 2}$
terms by parts, and substitute expressions for $a_x''$, $h_{tx}''$, $\xi''$
from the corresponding fluctuation equations. We need to keep track of the boundary
terms. Then let us go to the momentum space. As a result we arrive at $\tilde I_f=B_f'$, where:
\begin{align}
B_f&=-\frac{L^2 e^{-\frac{\chi }{2}}}{2 u^3} \left(e^{\chi } h_{tx}(u,-\omega ) \left(u^3 A_t' a_x(u,\omega )-u h_{tx}'(u,\omega )+4 h_{tx}(u,\omega )\right)\right.\notag\\
&+\left.u f \left(u^2 a_x(u,-\omega ) a_x'(u,\omega )+2 L^2 m^2 \xi (u,-\omega ) \xi '(u,\omega ) \dot V\right)\right)\,,
\label{Ifonsh}
\end{align}
where prime, as before, stands for a derivative w.r.t. $u$.

The counter-term action (\ref{ctact}) for the ansatz (\ref{flans}) is given by:
\begin{align}
&I^f_{c.t}=\frac{e^{-\frac{\chi }{2}}}{2 u^3 f^{1/2}} \left(2 L^4 m^2 u^2 f^2 \dot V\xi '(u,\omega )  \xi '(u,-\omega )+
L^2 e^{\chi } \left(h_{tx}(u,-\omega ) h_{tx}(u,\omega ) \left(2 L^2 m^2 V\right.\right.\right.\notag\\
&{-}\left.\left.\left. \psi ^2{-}4\right){-}2 L^2 m^2 u^2 
\dot V \left(\omega ^2 \xi (u,{-}\omega ) \xi (u,\omega ){+}\alpha  h_{tx}(u,{-}\omega ) (\alpha  h_{tx}(u,\omega ){+}2 i \omega  \xi (u,\omega ))\right)\right)\right)\,,\label{ctfl}
\end{align}
evaluated at $u=0$.

Now let us evaluate (\ref{ctfl}) minus (\ref{Ifonsh}) at $u=0$, which gives $I_{tot}^f$.
Consider the case $\Delta=2$. First we need to solve fluctuation
equations near the boundary.
We already determined the near-boundary asymptotics (\ref{nbbas}) for the background fields.
In superconducting phase we have $\psi_1=0$. Besides, from the equations of motion,
one obtains:
\beq
\gamma_1=-V_1\,m^2\,L^2\,\alpha^2\,.
\eeq
Similarly, the fluctuation equations of motion,
near the boundary give:
\begin{align}
\xi(u,\omega)&=\xi^{(1)}(\omega)+\xi^{(2)}(\omega)\,u^2+\xi^{(3)}(\omega)\,u^3\,,\notag\\
a_x(u)&=a_x^{(1)}+a_x^{(2)}\,u\,,\\
h_{tx}(u)&=h_{tx}^{(1)}(\omega)+h_{tx}^{(2)}(\omega)u^2+h_{tx}^{(3)}(\omega)\,u^3\,,
\end{align}
where again not all the coefficients of expansion are independent, and in fact:
\begin{align}
\xi^{(1)}(\omega)&=\frac{i}{2V_1m^2\alpha\rho\omega}\left(2\gamma_1\,a_x^{(2)}(\omega)+2V_1
m^2\alpha^2\rho\, h_{tx}^{(1)}(\omega)+\omega^2\,a_x^{(2)}(\omega)\right)\,,\label{xioneexpre}\\
\xi^{(2)}(\omega)&=\frac{i\omega}{4m^2V_1\,\alpha\rho}\left(2\gamma_1+\omega^2\right)\,a_x^{(2)}(\omega)\\
h_{tx}^{(2)}(\omega)&=\frac{1}{2\rho}\left(2\gamma_1+\omega^2\right)\,a_x^{(2)}(\omega)\\
h_{tx}^{(3)}(\omega)&=\frac{1}{3\omega}\left(\rho\omega\,a_x^{(1)}-6iV_1m^2\alpha\,\xi^{(3)}(\omega)\right)\,.
\end{align}
Using these asymptotic expansions, evaluating $I^f_{c.t}-B_f$ at $u=0$ gives\footnote{This is in agreement with eq. (3.14) of \cite{Kim:2015dna}.
See that only the leading linear term $V_1$
in the near-boundary expansion of $V(z)$ matters in this formula.} :
\begin{align}
I^f_{tot}&=a_x^{(1)}(-\omega)a_x^{(2)}(\omega)-\rho\,a_x^{(1)}h_{tx}^{(1)}(-\omega)
+2\gamma_2\,h_{tx}^{(1)}(-\omega)h_{tx}^{(1)}(\omega)-3h_{tx}^{(1)}(-\omega)h_{tx}^{(3)}(\omega)\notag\\
&+6m^2V_1\,\xi^{(1)}(-\omega)\xi^{(3)}(\omega)\,.
\end{align}
where we have kept $\xi^{(1)}$, for brevity (but keep in mind it is not an independent expansion coefficient,
due to (\ref{xioneexpre})).

It is convenient to replace $\xi\rightarrow Z$, so that we are dealing
with two fields, $(a_x\,,\,Z)$, which have the same near-boundary expansion,
at least up to the first two orders. Due to (\ref{zetaxi}), we obtain:
\beq
Z(u,\omega)=\frac{f(u)}{i\omega\alpha u}\xi'(u,\omega)\,,
\eeq
which near the boundary becomes:
\beq
Z(u,\omega)=Z^{(1)}(\omega)+Z^{(2)}(\omega)\,u+\cdots=-\frac{2i}{\alpha\omega}\,\xi^{(2)}(\omega)
-\frac{3i}{\alpha\omega}\,\xi^{(3)}(\omega)\,u+\dots\,.
\eeq

We can represent $I^f_{tot}$ in the form, convenient for calculation of correlation matrix:
\beq
I_{tot}^f=\left(a_x^{(1)}(-\omega),\, Z^{(1)}(-\omega)\right)\,{\cal M}\,
\left({a_x^{(2)}(-\omega)\atop Z^{(2)}(-\omega)}\right)+\cdots\,,
\eeq
where dots denote $\xi^{(1)}$ terms. We cannot extract $\xi^{(1)}$
by solving system of equations for $(a_x,\,Z)$, because $Z\sim\xi'$.
So we assume that $\xi^{(1)}$ is a constant of integration, which we fix to be:
\beq
\xi^{(1)}(\omega)=\frac{i(1+\sqrt{2})(\omega^2-2m^2\alpha^2V_1)}{2m^2\alpha\rho\omega V_1}\,a_x^{(2)}(\omega)\,,
\eeq
which is the choice enabling a diagonal matrix $M$.
Let us rescale the fluctuation fields (this is a symmetry transformation of fluctuation equations):
\beq
\left({a_x\atop Z}\right)\rightarrow \frac{1}{\sqrt{1-2\sqrt{2}+\frac{\sqrt{2}\omega^2}{m^2\alpha^2V_1}}}
\left({a_x\atop Z}\right)
\eeq
The corresponding matrix is:
\beq
{\cal M}=\left({1\atop 0}\;
{0\atop
 \frac{2m^2\alpha^2V_1}{\sqrt{1-2\sqrt{2}+\frac{\sqrt{2}\omega^2}{m^2\alpha^2V_1}}}}\right)\,.\label{MMdef}
\eeq
For the purpose of finding AC conductivity we only need the $(a_x,a_x)$ component of
the correlation matrix.

\end{document}